%% file: itrev3.tex
\documentclass[10pt,journal,final,letterpaper,twocolumn,twoside]{IEEEtran}

\IEEEoverridecommandlockouts  

\usepackage{times,latexsym,amsfonts,amsmath,amssymb,mathrsfs}
\usepackage{verbatim,cite}
\usepackage{epsfig,epsf}
\usepackage{floatflt}
\usepackage{graphicx}
\usepackage[usenames,dvips]{color}
\definecolor{myblue}{rgb}{0.2,0.4,1.00}

\def\nN{{\mathbb N}}
\def\zZ{{\mathbb Z}}
\def\rR{{\mathbb R}}
\def\eE{{\mathbb E}}
\def\pP{{\mathbb P}}

\def\QED{\mbox{\rule[0pt]{1.5ex}{1.5ex}}}

\def\@begintheorem#1#2{\tmpitemindent\itemindent\topsep 0pt\rm\trivlist
    \item[\hskip \labelsep{\indent\it #1\ #2:}]\itemindent\tmpitemindent}
\def\@opargbegintheorem#1#2#3{\tmpitemindent\itemindent\topsep 0pt\rm \trivlist
    \item[\hskip\labelsep{\indent\it #1\ #2\
    \rm(#3):}]\itemindent\tmpitemindent}
\def\@endtheorem{\endtrivlist\unskip}

\newtheorem{lemma}{Lemma}[section]
\newtheorem{fact}{Fact}[section]
\newtheorem{proposition}{Proposition}[section]
\newtheorem{definition}{Definition}[section]
\newtheorem{remark}{Remark}[section]

\renewcommand{\theequation}{\arabic{section}.\arabic{equation}}
\newcommand{\Section}[1]{\section{#1}
\setcounter{equation}{0}}

\title{High-resolution distributed sampling of bandlimited fields with
  low-precision sensors$^*$\thanks{$^*$ An early preliminary part of
  this work was presented in IPSN and Allerton 2003. This material is
  based upon work supported in part by the US National Science
  Foundation (NSF) under awards (CAREER) CCF--0546598, (SENSORS)
  CCR-0330514, CCR-0219722, and DARPA F30602-00-2-0538. Any opinions,
  findings, and conclusions or recommendations expressed in this
  material are those of the authors and do not necessarily reflect the
  views of the NSF or DARPA.}}

\author{Animesh Kumar, Prakash Ishwar, and Kannan
Ramchandran$^\dagger$\thanks{$^\dagger$~P.~Ishwar is with the ECE
Dept.,~Boston University, Boston, MA 02215. A.~Kumar and
K.~Ramchandran are with the EECS Dept.,~Univ.~California, Berkeley, CA
947210. Emails: {\tt\small pi@bu.edu}, {\tt\small \{animesh,kannanr\}@eecs.berkeley.edu}.}
}

\begin{document}

\maketitle 

\thispagestyle{plain}
\pagestyle{plain}

\begin{abstract}
The problem of sampling a discrete-time sequence of spatially
bandlimited fields with a bounded dynamic range, in a distributed,
communication-constrained, processing environment is addressed.  A
central unit, having access to the data gathered by a dense network of
low-precision sensors, operating under stringent inter-node
communication constraints, is required to reconstruct the field
snapshots to maximum accuracy. Both deterministic and stochastic field
models are considered. For stochastic fields, results are established
in the almost-\-sure sense.  The feasibility of having a flexible
tradeoff between the oversampling rate (sensor density) and the
analog-to-digital converter (ADC) precision, while achieving an
exponential accuracy in the number of bits per Nyquist-interval
{per snapshot} is demonstrated. This exposes an underlying
{\em ``conservation of bits'' principle}: the bit-budget per
Nyquist-interval per snapshot (the rate) can be distributed along the
amplitude axis (sensor-precision) and space (sensor density) in an
almost arbitrary discrete-valued manner, while retaining the same
(exponential) distortion-rate characteristics. An achievable
information scaling law for field reconstruction over a bounded region
is also derived. With $N$ one-bit sensors per Nyquist-interval,
$\Theta(\log N)$ Nyquist-intervals, and total network bitrate $R_{net}
= \Theta((\log N)^2)$ (per-sensor bitrate $\Theta((\log N)/N)$), the
maximum pointwise distortion goes to zero as $D = O((\log N)^2/N)$ or
$D = O(R_{net} 2^{-\beta \sqrt{R_{net}}})$. This is shown to be
possible with only nearest-neighbor communication, distributed coding,
and appropriate interpolation algorithms.  For a fixed, {\it nonzero}
target distortion, the number of sensors and the network rate needed
is always {\it finite}.
\end{abstract}


\Section{Introduction}

{\it Motivation:} High resolution remote sensing of physical phenomena
is a task of considerable importance in applications such as
environment/weather monitoring, ecology, and precision
agriculture. Consider the scenario where a large distributed network
of low-precision, low-power sensors is deployed over a region of
interest to collect and return measurements to a central data
collection and processing unit (CPU)\footnote{Sensors can take turns
to be the CPU.}  over some time duration. The goal is to efficiently
acquire, process, and transport data to faithfully reconstruct the
physical field under appropriate measures of fidelity.

Physical fields such as temperature, pressure, magnetization, and
vibration are fundamentally analog in nature. Their spatio-temporal
distributions obey fundamental physical laws (a temperature field must
satisfy Laplace's equation, the velocity field in fluid flows must
satisfy Navier-Stokes equation, etc.) that induce strong dependencies
across space and time. Many physical fields are approximately
bandlimited { in space and time} {(the physical
propagation laws often providing a natural filtering effect that
attenuates high frequencies)} and can be reconstructed in a stable
manner from samples taken slightly above the Nyquist-rate on a uniform
lattice.  However, in practice, the samples of the field are quantized
due to the finite precision of ADCs, leading to unavoidable field
reconstruction errors. When bandlimited fields are uniformly sampled
at the critical Nyquist-rate {(in space and time)}, the
worst-case pointwise reconstruction error $D(R)$ decays exponentially
with the bitrate $R$ (measured in number of bits per Nyquist-interval
per snapshot{ \footnote{{Sensors are assumed to
periodically sample the field in synchronism at discrete time instants
at their locations. A field snapshot refers to the entire
continuous-space field at one sampling time instant.}}}  of the ADCs
\cite{marksi1990,higginss1996,ThaoV-RO:TransSP94,ThaoV-DA:TransSP94,ZoranV-ER:IT98},
that is, $D(R)$ is of the order of $O(2^{-R})$.
Error can be reduced { by increasing the quantization
resolution of the} sensors.
Is it possible to sacrifice ADC-precision (lower amplitude resolution)
for denser spatial sampling rate (higher spatial resolution) while
maintaining the same reconstruction quality and the same number of
information bits per Nyquist-interval {per snapshot}? In
the sequel we show that it is indeed possible to achieve a flexible
array of tradeoffs between the (amplitude) resolution and the density
(spatial resolution) of sensors leading to a {\em ``conservation of
  bits''} principle.

The classical problem of sampling of continuous signals is a mature
topic in signal processing that has accumulated a rich knowledge-base
over the past several decades
\cite{marksi1990,higginss1996,Marvasti-NS:Plenum01}. However, sensor
networks impose challenging constraints on the classical sampling
paradigm in terms of low device precision and power and prohibitive
communication costs. This biases solutions toward highly distributed
and/or localized collaborative processing environments characterized
by nearest-neighbor communication. This also calls for distributed
compression algorithms to be integrated with sampling. We use the term
``distributed sampling'' to capture these effects.  This paper is
accordingly driven by the goal of addressing the sampling problem
within the context of sensor networks and their associated constraints
in terms of both localized processing and communication. At an
information processing level, there are two broad functional tasks in
data-gathering sensor networks: (i) data-acquisition or sampling,
processing, and coding of the sensor field measurements and (ii) data
transport, where the acquired data is disseminated across the network
to a CPU.  Both tasks are addressed in this work.  The
data-acquisition task is addressed through a distributed sampling
framework that is well-suited for sampling sensor fields with minimal
or no inter-sensor communication.  The data-transport task is
addressed through distributed coding and nearest-neighbor
communication that moves the sensor data to the CPU. For a related
information-theoretic study of multiuser source coding with
distributed 
vector quantization, { see
\cite{KashyapLXL:DCC05,NeuhoffP-2006-UPRIDLSC}. For work with a
different flavor on quantization and oversampling, see
\cite{marcodlnoIPSN2003}.}

{\it Contributions:} (i) {\it ``Bit-conservation'' principle:} A key
contribution of this work is that the bit-budget per Nyquist-interval
(the rate) can be distributed along the amplitude axis
(sensor-precision) and space (sensor density) in an almost arbitrary
discrete-valued manner, while retaining (the best-known) exponential
distortion-rate characteristics (Section~\ref{sec:bcp}).  This
provides the network-\-architect with a guiding design principle for
{\em spatially adaptive sampling} in terms of selecting sensor
precision, density, and deployment patterns to meet the desired
reconstruction quality. We can use higher precision ADCs when the
sampling density is constrained to be light by the underlying
topography of sensor-deployment, e.g., rugged terrain or occluding
obstacles, and use lower precision sensors when the sampling density
can be high. This also clarifies the need for {\em diversity} in
sensing capabilities for easing the adaptation to changing
environments.

(ii) {\it Information scaling law:} The pioneering work of Gupta and
Kumar \cite{guptaktcIT2000} that has precipitated interest in scaling
laws for ad hoc networks, applies to scenarios where nodes produce
{\em independent data}, no matter what the scale of the network.  This
does not entirely apply to the sensor network context, where the
``information density'' (related to the sensed field) remains fixed
regardless of the ``network density'' (related to the network size).
The inter-sensor data correlation induced by the underlying physics of
the phenomenon being sensed can be exploited to reduce the net
information to be disseminated in the network. How does network
information grow with increasing demands on quality?  What are the
network resources needed to sustain high-quality large-scale field
acquisition?  How should the underlying spatio-temporal correlation be
exploited?  For reconstructing a discrete-time sequence of spatially
bandlimited fields over a fixed bounded region at a CPU, we show that
the following scaling behavior can be realized with one-bit sensors,
distributed coding, neighbor-to-neighbor local communication, and
suitable spatial interpolation (Section~\ref{sec:DataTran}): With $N$
one-bit sensors per Nyquist-interval, $\Theta(\log N)$
Nyquist-intervals, and total network bitrate $R_{net} = \Theta((\log
N)^2)$ (per-sensor bitrate $\Theta((\log N)/N)$), the maximum
pointwise distortion goes to zero as $D = O((\log N)^2/N)$ or $D =
O(R_{net} 2^{-\beta \sqrt{R_{net}}})$.

(iii) {\it Stochastic fields:} An important technical contribution of
this work is the generalization of the nonuniform sampling results for
amplitude limited, bandlimited, square-integrable deterministic fields
due to Cvetkovi{\'c} and Daubechies \cite{zorands2000} to amplitude
limited, bandlimited, wide-sense stationary (WSS) spatial stochastic
processes (Sections~\ref{sec:keytech}, \ref{sec:NQstochext},
\ref{sec:onebitstochext}, and \ref{sec:bcpstochext}).  This
generalization is effected by leveraging certain results for the Zakai
class of bandlimited fields due to Zakai \cite{ZakaiInfoCntrl65} and
Cambanis and Masry \cite{CambanisMasrySIAM76}. The maximum pointwise
error versus rate results are established in the strong almost-\-sure
sense (hence also in the expected $p$-th power sense, $p \in
(0,\infty)$). We build upon the work by Cvetkovi{\'c} and Daubechies
along three directions. First, we generalize their results to
arbitrary precision ADCs. Secondly, we extend their framework to the
stochastic setting.  Thirdly, we leverage these results from the
classical centralized setup to the distributed setting accounting for
information transport costs making them relevant to the sensor network
context.

\noindent{\em Key ideas:} A key concept underlying the ability to
achieve a flexible tradeoff between amplitude and spatial resolution
is dithered sampling { due to Cvetkovi{\'c} and Daubechies
\cite{zorands2000}}. For each field snapshot, sensors add (or compare)
the value of a pre-designed dither function at their respective
locations to their observations and note only the {\em sign} of the
sum. By design of the dither function, the {\em sum} of the
bandlimited field and the dither function is ensured to have {\em
exactly one zero-crossing in every Nyquist-interval}. This induces a
strong dependency in the binary observations of all $N = 2^R$ sensors
inside each Nyquist-interval. Their joint entropy-rate is ensured to
be not more than $R$ bits per snapshot. This data can be compressed
using a distributed source code and moved to the CPU using
nearest-neighbor communication. The total rate for this is not more
than $R$ bits per Nyquist-interval per snapshot. The {\em spatial
resolution} of the { reconstructed zero-crossings}, which
is proportional to inter-sensor separation $= O(2^{-R})$, can be
translated to {\em amplitude resolution} of the {
reconstructed} field at these locations through the local smoothness
properties of the field and the dither (mean-value theorem). Using
results in nonharmonic Fourier analysis { due to
Cvetkovi{\'c} and Daubechies \cite{zorands2000}}, per-sample amplitude
resolution can be converted, { as will be shown}, to
global field approximation accuracy of the same order using
interpolation techniques. The maximum reconstruction error using $2^R$
one-bit sensors per Nyquist-interval { will then be of}
the order of $O(2^{-R})$. This is equivalent to the reconstruction
accuracy of one $R$-bit sensor per Nyquist-interval.  A similar
approach using dithered {\em level}-crossings instead of
zero-crossings allows one to increase precision and decrease density
maintaining order-optimal accuracy.

{\it Caveats:} This work has a sampling-theoretic focus.
{ Although the motivation is two-dimensional sensor
networks, for clarity, the theoretical development focuses on a single
time-snapshot and spatial dimension equal to one. The multidimensional
case is discussed in the last section.} We { also}
acknowledge that there are several important problems that must be
overcome to make the results of this work practically viable. This
includes sensing-noise, sensor-deployment issues, sensor location
errors or uncertainties, and sensor synchronization and scheduling
issues. We defer the exploration of such issues to future research.

{\it Organization:} Field modeling assumptions, distortion criteria,
and assumptions on data-acquisition and communication, are discussed
in Section~\ref{sec:models}.  Section~\ref{sec:keytech} discusses the
key technical issues and results in going from the deterministic to
the stochastic setup.  Throughout, the deterministic and the
stochastic scenarios are discussed side-by-side to emphasize
similarities and differences.  Classical deterministic Nyquist
sampling and its stochastic extension is discussed in
Section~\ref{sec:NQintro}. Section~\ref{sec:ditherintro} discusses
single-bit dithered oversampling, first by summarizing the
deterministic results by Cvetkovi{\'c} and Daubechies
\cite{zorands2000} and then extending it to the stochastic case. A
{\it ``bit-conservation''} principle and its implications are exposed
in Section~\ref{sec:bcp} by showing how a flexible trade-off between
sensor-precision and sensor density can be achieved.  {
Distributed processing and communication issues and costs germane to
sensor networks are discussed in Section~\ref{sec:sensornetworks}.
This includes distributed field-acquisition and coding
(Section~\ref{sec:DataAcqu}), information transport and associated
achievable information scaling law in terms of the bitrate,
distortion, and sensor density for field reconstruction in a compact
region of interest (Section~\ref{sec:DataTran}), and extensions to two
and higher spatial dimensions (Section~\ref{sec:2Dextn}).} The main
contributions are summarized in Section~\ref{sec:conclusion}.  Proofs
of technical results are relegated to the appendices to maintain a
smooth flow of ideas.

{\it Notation:} Real numbers, the integers, and the natural numbers
are respectively denoted by $\rR$, $\zZ$, and $\nN$.  The term `field'
is used to emphasize the dependence of a quantity on {\em spatial}
coordinates, e.g., electromagnetic field, temperature field, pressure
field, etc.  All random variables and processes are defined with
respect to a common probability space $(\Omega,\mathcal{F},\pP)$ and
$\eE[\cdot]$ denotes the mathematical expectation operator. Random
quantities are denoted by capital letters (e.g., $X$) and specific
realizations of random quantities by small letters (e.g., $x$).
Modifications of $f(t)$ and $X(t)$, e.g., $f', X', \widehat{f},
\widehat{X}$, etc., respectively denote deterministic and stochastic
fields and $t, T, t_l, T_l$, etc., are used for spatial (not temporal)
variables. The notation $X$ without adornments denotes the {\it
entire} process $\{X(t)\}_{t\in\rR}$. $\mathcal{L}^p(\rR^d)$ denotes
the space of $p$-th power ($p \in (0, \infty)$), Lebesgue-integrable
fields on $\rR^d$ where $d =$ dimension. $BL(S)$ denotes the space of
deterministic, square-\-integrable (finite-energy), continuous,
bandlimited fields\footnote{\label{fn:cont} In $\mathcal{L}^2(\rR^d)$
a bandlimited field can be discontinuous on a measure-zero set that
may even be uncountably infinite. The continuity condition is
explicitly included due to our interest in sample-interpolated
reconstructions.} on $\rR^d$, with a compact spectral support set
$S$. The phrase `almost-\-surely' is synonymous with the phrase `with
probability one'. The term bitrate is used to denote bits per spatial
Nyquist-interval/area/volume for each field snapshot.  We follow
Landau's asymptotic notation: $f(x) = O(g(x))$ as $x \rightarrow a
\Leftrightarrow \lim\sup_{x\rightarrow a}|f(x)/g(x)| < \infty$; 
$f(x) = \Theta(g(x)) \Leftrightarrow f(x) = O(g(x))\ \text{and}\ g(x)
= O(f(x))$. { The value of $a$ is either $\infty$ or $0$
and will always be clear from the context.}

\Section{Modeling Assumptions and Fidelity Criteria \label{sec:models}}

We consider a discrete-time sequence of continuous-space spatially
bandlimited fields with a limited amplitude range. Both deterministic
and stochastic field models, described below in
Section~\ref{sec:fieldmodel}, are studied. These modeling assumptions
hold for each discrete-time field snapshot with spatial dimension $d$
equal to one. An extension to higher spatial dimensions is discussed
in Section~\ref{sec:sensornetworks}. When dependencies across time
snapshots are arbitrary, our error versus rate results apply only to
each snapshot and not to other intermediate time instants.  If the
field is temporally bandlimited, snapshots are assumed to be taken at
the temporal Nyquist-rate.
Then our results apply to all spatio-temporal points.  Knowledge of
the field model and associated parameters such as bandwidth and
dynamic range are assumed to be available during the design-phase
prior to sensor deployment. Sensors are assumed to be deployed in a
uniform rectangular grid and the sensor locations are assumed to be
available to the fusion center.
Sensors are assumed to operate in a time-synchronous manner.
Although joint source-channel issues and power-distortion considerations
are not considered in this work, the data-transport costs in terms of
bitrate are studied in Section~\ref{sec:sensornetworks}. The
finite-rate encoded sensor observations together with their sensor
identification labels are made available to a fusion center.  The
objective of the fusion center is to estimate the field value at each
continuous-space location for each discrete-time snapshot. The
estimation quality is measured by the maximum absolute error over all
space and time. Bounds on these maximum errors over all space and time
also imply bounds on related space and time {\em averaged} distortion
criteria described in Section~\ref{sec:distcrit}.

\subsection{Field-models \label{sec:fieldmodel}}

The following modeling assumptions apply to {\it each} temporal
snapshot of the field that is acquired by the sensor network. For
clarity, the theoretical development focuses on a single time-snapshot
and spatial dimension equal to one.  The multidimensional and temporal
aspects are discussed in Section~\ref{sec:sensornetworks}.

\subsubsection{Deterministic case} 

We assume that $f(t),\ t\in \rR$, is an amplitude limited field, with
amplitude limit $A$, belonging to $BL([-W,W])$. Here, both $A$ and $W$
are some finite, strictly positive, real numbers.  Distributed
sampling of deterministic, square-\-integrable, {\em non-bandlimited}
fields, with an exponentially decaying spectrum, can also be studied
using the techniques discussed here (see \cite{kumariroICASSP2004} for
details), but will not be entered into.
It should be noted that a limited amplitude constraint is less
restrictive than a limited energy constraint for bandlimited fields. A
bandlimited field with limited energy is necessarily amplitude limited
but the converse need not be true, e.g., $\forall t \in \rR$, $f(t) =
\sin(t)$. The modeling assumptions admit fields with an arbitrarily
high but finite energy.

A property of bandlimited fields that plays an important role in the
derivation of field-interpolation error bounds from nonuniform samples
\cite{zorands2000}, and used in our work, is the uniformly bounded
slope property.  Specifically, according to Bernstein's
inequality~\cite{hardylpi1959}, \cite[p.~144]{PapoulisSigAnal}, for
fields in $BL([-W,W])$ with amplitude limit $A$,
\begin{eqnarray}
|f'(t)| \leq W \sup_{t\in \rR}|f(t)| \leq W A,\quad \forall t\in\rR.
\label{eq:Bernstein}
\end{eqnarray}
Through a suitable renormalization of the amplitude and spatial axes,
it can be assumed, without loss of generality, that the amplitude
ranges of $f$ and $f'$ are respectively contained in $[-1,1]$ and
$[-\pi,\pi]$ and the spectral support of $f$ is contained in
$[-\pi,\pi]$. This corresponds to $W = \pi$ and $A = 1$.

\subsubsection{Stochastic case} \label{sec:stochfldmdl} 

We assume that $X(t), t\in \rR$, is an amplitude limited stochastic
process (field), with amplitude limit $A$, which is WSS with an
autocorrelation function $R_X(t), t \in \rR$, belonging to
$BL([-W,W])$. Note that the process $X$ is mean-square continuous
since $R_X$ is continuous.\footnote{Continuity at $t=0$ suffices since
it implies continuity $\forall t \neq 0$ \cite{papoulisp1965}.}
Although somewhat restrictive, the class of processes which are
stationary, bandlimited, and amplitude limited, is quite rich. Some
examples are presented in Appendix~\ref{ap:WSSconstr}.

\subsection{Distortion criteria \label{sec:distcrit}}

The performance of different sampling and reconstruction schemes will
be compared by deriving a rate-dependent upper bound for the maximum
(over space) field reconstruction error, that is, the
$\mathcal{L}^{\infty}$-norm of the error.

\subsubsection{Deterministic case} 

Let $\widehat{f}_R(t)$ denote the reconstruction of {a
  single field snapshot} $f(t)$ at the spatial location $t \in \rR$,
according to some sampling, quantization, and reconstruction scheme
{which uses} $R$ bits per meter per snapshot. Let $e(t) =
(f(t) - \widehat{f}_R(t))$ denote the reconstruction error at $t$.
For different sampling and reconstruction schemes, we
{will} derive upper bounds $\eta(R)$, on the maximum (over
space $t$) field reconstruction error, $D(R) := ||e||_\infty$.  These
bounds then automatically hold for the so-called {\em
  locally-averaged}\footnote{This is a deterministic spatial average
  and not a stochastic average with respect to any probability
  distribution.}  $\mathcal{L}^q$ quasi-norm\footnote{Note that this
  is a norm only for $ q \in [1,\infty)$. We do not require this to be
    a norm so we allow $q \in (0,\infty)$.} of the reconstruction
  error $e$ defined by
\begin{eqnarray*}
||e||_{q}(t,T) &:=& \left\{\frac{1}{T}\int_{|s-t|<T/2}|e(s)|^q{\rm
d}s\right\}^{\frac{1}{q}},\quad q \in (0,\infty),
\end{eqnarray*}
because for all $t \in \rR$ and all $T \in (0,\infty)$,
$||e||_{q}(t,T) \leq ||e||_{\infty}$.

\subsubsection{Stochastic case}
{ Let $\widehat{X}_R$ and $E = (X - \widehat{X}_R)$
  respectively denote the reconstruction and error fields according to
  some sampling, quantization, and reconstruction scheme which uses
  $R$ bits per meter per snapshot.  It should be noted that $E$ may
  not be WSS because the reconstruction scheme may be nonlinear (due
  to quantization) and spatially varying (due to sampling). For
  different sampling and reconstruction schemes, we will derive upper
  bounds $\eta(R)$ such that $||E||_\infty \leq \eta(R)$ almost surely
  and consequently also $||(\eE|E|^p)^{1/p}||_\infty \leq \eta(R)$ for
  any $p \in (0,\infty)$.  As in the deterministic case, these bounds
  will also apply to the locally-averaged error
  $\left(||E||_{q}(t,T)\right)$ both in the almost-\-sure and expected
  $p$-th power senses.}

\Section{From Deterministic to Stochastic Fields} \label{sec:keytech}

To expose the intuition underlying the main results of this work and
for expositional clarity, in the sequel, we first discuss results for
the deterministic setting, and then immediately extend this to the
stochastic setting.  Moving from the deterministic to the stochastic
setting would have been straightforward if almost all sample paths of
$X$ inherited the same properties assumed for deterministic fields.
{ The deterministic nonuniform sampling and interpolation
results for amplitude and bandlimited fields by Cvetkovi\'{c} and
Daubechies
\cite{zorands2000}, that we build upon, are stated for
square-\-integrable fields that also satisfy a crucial uniformly
bounded slope property (Bernstein's inequality
(\ref{eq:Bernstein}). Moreover, the nonuniform sampling locations in
these deterministic results, are intricately coupled to field values,
because they are derived from certain level-crossings of the field.
It is unclear if these conditions hold in the stochastic setting. For
instance,} while every sample path of $X$ is amplitude limited, it
need not be bandlimited in the conventional sense.  In fact, the
sample paths need not even be square-\-integrable,\footnote{If $X(t)$
is a strictly stationary ergodic process with $0 < R_X(0)$, then
almost-\-surely $ R(0) = \lim_{T\rightarrow \infty}
(2T)^{-1}\int_{[-T,T]}X^2(t)\mbox{d}t$. Thus almost all sample paths
are not square-integrable.} which is a common assumption in results
pertaining to the approximation of deterministic fields from a
discrete set of samples.  The extension of sampling results from the
conventional square-\-integrable bandlimited fields to the stochastic
case is one of the key contributions of our work.

It turns out that almost all sample paths of a mean-square continuous,
WSS stochastic process, with an autocorrelation function in
$BL([-W,W])$, are bandlimited in the sense of
Zakai~\cite{ZakaiInfoCntrl65,CambanisMasrySIAM76,BleyaevSIAM61}. This
is an extension of the conventional notion of bandlimited fields.  The
Zakai class of bandlimited fields (defined below) includes the
conventional square-\-integrable bandlimited fields but also includes
fields such as nonzero constants and pure tones (sinusoids) which are
not square-\-integrable. Therefore, our approach for extending
sampling results from the deterministic to the stochastic setting is
to first establish deterministic sampling results for the Zakai class
of bandlimited fields, which includes non square-\-integrable fields,
via Lemma~\ref{lemma:Zakaiextn}. The results for the stochastic
setting follow immediately, in a strong almost-\-sure sense, via
Fact~\ref{fact:samplepaths} below. The results then automatically hold
in an expected $p$-th power sense as well.

\begin{definition}({\em Zakai class of bandlimited
  fields~\cite{ZakaiInfoCntrl65,CambanisMasrySIAM76,BleyaevSIAM61}
  $ZBL(W,\delta)$}) \label{def:zakaiclass} { A field $f(t)$, $t\in
  \rR$, is said to be bandlimited in the sense of Zakai, with
  bandwidth $W$ and a margin parameter $\delta \in (0,\infty)$, if
  $\int_{-\infty}^{\infty} |f(t)|^2 \frac{1}{1+t^2}\mbox{d}t < \infty$
  and
\begin{eqnarray}
f(t) = \int_{-\infty}^{\infty} f(\tau) h(t-\tau) \mbox{d}\tau,
\label{eq:reprodK}
\end{eqnarray}
where $h(t) = \frac{2}{\pi \delta t^2} \sin((W + \delta/2)t)
\sin(\delta t /2)$ for $t\neq 0$ and $h(0) :=
(2\pi)^{-1}(2W+\delta)$. {$ZBL(W,\delta)$ is the set of
  all fields which are bandlimited in the sense of Zakai with
  bandwidth $W$ and margin parameter $\delta$.}}
\end{definition}

\begin{figure}[!htb]
\begin{center}
\scalebox{0.4}{\input{./Figures/Zakai.pstex_t}}
\end{center}
\caption{\label{fig:Zakai} \sl \small {\bf Fourier transform of kernel
    $h(t)$.} }
\end{figure}
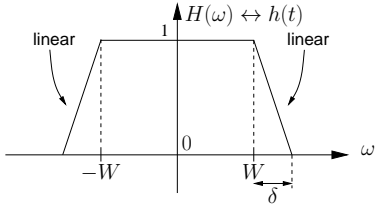

To provide some insight into the Zakai class of bandlimited fields, we
summarize some of its salient properties in this paragraph. These
properties are not needed in the subsequent development.  The function
$h$ appearing in the definition of $ZBL(W,\delta)$ has a bounded
amplitude and is both absolutely and square integrable because it
decays sufficiently fast (as $1/t^2$). It has a Fourier transform
given by the trapezoidal-shaped function shown in
Figure~\ref{fig:Zakai}.
Thus $h \in BL([-W-\delta,W+\delta])$. The set of all Borel-measurable
complex-valued functions $f(t)$ on the real line satisfying
$\int_{-\infty}^{\infty}|f(t)|^2 \frac{1}{1+t^2} \mbox{d}t < \infty$
forms a Hilbert space with inner product $<f,g> =
\int_{-\infty}^{\infty} f(t)g^*(t) \frac{1}{1+t^2}\mbox{d}t$ where
$^*$ denotes complex conjugate. This Hilbert space includes
$\mathcal{L}^2(\rR)$, bounded functions, and functions for which
$\frac{1}{2T} \int_{-T}^T|f(t)|^2\mbox{d}t$ is bounded in $T$, e.g.,
$f(t) = \sin(t),~t\in\rR$ \cite{ZakaiInfoCntrl65}. $ZBL(W,\delta)$
then forms a reproducing-kernel subspace of this Hilbert space with
the Toeplitz reproducing kernel $h$. Every function in $ZBL(W,\delta)$
is equal almost everywhere (with respect to the Lebesgue measure) to a
continuous function (given by the right side of (\ref{eq:reprodK}))
and, as in the deterministic case (see footnote~\ref{fn:cont}), only
these continuous representatives will be considered. For all $\delta,
W$, $BL([-W,W]) \subseteq ZBL(W,\delta)$, that is, conventionally
(square-\-integrable) bandlimited fields are also Zakai-sense
bandlimited. It is possible to use other kernels for $h$, e.g.,
kernels whose Fourier transforms increase and decrease super-linearly
over $[-W-\delta,-W]$ and $[W,W+\delta]$ respectively. Extending prior
work due to Zakai \cite{ZakaiInfoCntrl65}, in
\cite{CambanisMasrySIAM76} Cambanis and Masry showed that for all $W,
\delta$, a field $g \in ZBL(W,\delta)$ if and only if, for all $t\in
\rR$, $g(t) = g(0) + t f(t)$, for some $f \in BL([-W,W])$. Thus,
$ZBL(W,\delta)$ is in fact independent of the value of the margin
parameter $\delta$ which can be taken to be any arbitrarily small but
strictly positive real number and mainly serves to simplify
convergence issues.

The following uniformly bounded slope property for amplitude limited
fields in $ZBL(W,\delta)$, proved in Appendix~\ref{ap:bddslope}, is
crucial for establishing the asymptotic exponential decay of
approximation error with bitrate for the stochastic case.  This is
discussed in Sections~\ref{sec:ditherintro} and~\ref{sec:bcp} in the
context of sampling using low-precision ADCs.  The bound provided is
sufficient for our purpose although it could potentially be
strengthened to something more like Bernstein's inequality
(\ref{eq:Bernstein}).

\begin{proposition}({\em Amplitude limited fields in $ZBL(W,\delta)$ have
  bounded slope}) \label{prop:bddslope} {Let $0 < W, \delta, A <
\infty$, $f \in ZBL(W,\delta)$ continuous, $|f(t)| \leq A$ for all
$t\in \rR$, and $h$ be as in Definition~\ref{def:zakaiclass}. Then $f$
is differentiable and { $|f'(t)| \leq 2AW^2 < \infty$}
for all $t\in\rR$.}
\end{proposition}

The following key result, proved in Appendix~\ref{ap:Zakaiextn}, shows
that general uniform/nonuniform sampling and interpolation results
which hold for amplitude limited fields in $BL([-W,W])$, will also
continue to hold for amplitude limited fields in $ZBL(W,\delta)$, for
all $W,\delta$.

\begin{lemma}({\em Lifting sampling and interpolation results from
  $BL([-W,W])$ to $ZBL(W,\delta)$}) \label{lemma:Zakaiextn} {Let $0 <
  W, \delta, A < \infty$ and $h \in BL([-W-\delta,W+\delta])$ be the
  Toeplitz reproducing kernel in Definition~\ref{def:zakaiclass}. Let
  $\mathcal{T} = \{t_l\}_{l\in\zZ}$, a set of sampling locations, and
  $\{\psi_l\}_{l\in\zZ}$ a set of interpolation kernels be such that
  (i) $C := \sup_{t\in\rR}\left[\sum_{l\in\zZ}|\psi_l(t-t_l)|\right] <
  \infty$ and (ii) for all $(t,\tau) \in \rR^2$,
\begin{eqnarray*}
h(t-\tau) = \lim_{L\longrightarrow\infty}\sum_{l=-L}^{L} h(t_l - \tau)
\psi_l(t-t_l)
\end{eqnarray*}
{(the existence of such $\mathcal{T}$ and
$\{\psi_l\}_{l\in\zZ}$ is ensured by Fact~\ref{fact:zoransmainthm}).}
Then for any $f \in ZBL(W,\delta)$ with amplitude limit $A$,
\begin{eqnarray*}
f(t) &=& \lim_{L\rightarrow \infty}\sum_{l=-L}^L f(t_l)\psi_l(t-t_l),
\end{eqnarray*}
where the above series converges absolutely for each $t$ and uniformly
on all compact subsets of $\rR$.}
\end{lemma}

Sampling and interpolation results which hold for fields in
$ZBL(W,\delta)$, also hold true for the sample paths of a WSS process
with an autocorrelation function in $BL([-W,W])$, in the almost-\-sure
sense, due to the following fact.

\begin{fact}({\em Sample paths of mean-square continuous bandlimited WSS
  processes \cite[Sec.~5]{ZakaiInfoCntrl65},
  \cite[Sec.~3]{CambanisMasrySIAM76}})
\label{fact:samplepaths} { Let $0 < W, \delta < \infty$. 
If $X$ is a WSS stochastic process with an autocorrelation function
$R_X \in BL([-W,W])$, then almost-\-surely (with probability one), its
sample paths belong to $ZBL(W,\delta)$.}
\end{fact}

{ Thus, if there is an $R$ bits per meter per snapshot
  sampling, quantization, and reconstruction scheme such that
  $||f-\widehat{f}_R||_\infty \leq \eta(R)$ for all $f \in
  ZBL(W,\delta)$ then for any WSS $X$ with $R_X \in BL([-W,W])$,
  almost-\-surely, $||X(t) - \widehat{X}_R||_\infty \leq \eta(R)$.
  Thus an upper bound on the maximum error that holds for
  deterministic fields which are Zakai-sense bandlimited, also holds
  for bandlimited WSS stochastic fields in a strong almost-\-sure
  sense.}

As in the deterministic field model, through a suitable
renormalization of the amplitude and spatial axes, it can be assumed,
without loss of generality, that the amplitude ranges of $f$ and $f'$
are respectively contained in $[-1,1]$ and $[-\pi,\pi]$ and the
spectral support of $f$ is ``essentially'' contained in
$[-\pi,\pi]$. This corresponds to $W = \pi$ with $\delta$ arbitrarily
small but fixed, and { $A = 1/(2W^2)$.}

\Section{Nyquist-sampling (improving quality through sensor-precision)
\label{sec:NQintro}}

\subsection{Deterministic case}

Any field $f \in BL([-\pi,\pi])$ can be reconstructed from the samples
$\{f(l)\}_{l\in\zZ}$ according to the well known interpolation formula
\cite{marksi1990}
\begin{equation}
f(t) = \lim_{L\longrightarrow\infty}\sum_{l=-L}^{L} f(l)
\mbox{sinc}(t-l), \quad \forall t \in \rR,
\label{eq:classicinterp}
\end{equation}
where $\mbox{sinc}(t) := (1/\pi t)\sin(\pi t)$ for $t \neq 0$ and
$\mbox{sinc}(0) := 1$. The Nyquist-sampling period for $f$ is $T_{NQ}
= 1 $.  However, in practice the reconstruction
(\ref{eq:classicinterp}) is not stable to bounded perturbations in the
sample values due to the poor decay properties of the $\mbox{sinc}$
interpolation kernel, that is, the series in (\ref{eq:classicinterp})
is not absolutely-\-convergent. All practical ADCs have finite
precision and the sampling process is invariably subject to
perturbations. The instability in the reconstruction implies that the
quantization noise/error can potentially build up and lead to
unbounded reconstruction errors in parts of the
field.\footnote{However, if the perturbations are only due to scalar
quantization where $0$ is a reproduction point, then the interpolation
error will be bounded for fields in $BL([-\pi,\pi])$.}  However, the
instability can be overcome by taking samples slightly above the
Nyquist-rate:

\begin{fact}({\em Stable interpolation of bandlimited fields from samples
  at uniformly spaced locations \cite{zorands2000}})
\label{fact:stableinterp} { For each $\lambda > 1/T_{NQ} = 1$, there
exists an absolutely-\-integrable kernel $\phi_{\lambda}(t)$ belonging
to $BL([-\pi\lambda,\pi\lambda])$ such that $C :=
\sup_{t\in\rR}(\sum_{l\in\zZ} |\phi_{\lambda}(t-(l/\lambda))|) <
\infty$ and for all fields $f \in BL({[-\pi\lambda +
\delta,\pi\lambda - \delta]})$ with $0 < \delta <
{\pi\lambda}$,
\begin{equation}
f(t) = \lim_{L\longrightarrow\infty}\sum_{l=-L}^{L}
f(l/\lambda) \phi_{\lambda} \left(
t-\frac{l}{\lambda}\right), \quad \forall t \in \rR,
\label{eq:stableinterp}
\end{equation}
where the above series converges absolutely for each $t$ (and in
$\mathcal{L}^2$) and uniformly on all compact subsets of $\rR$.}
\end{fact}

It is due to the finiteness of $C$ that the reconstruction series
(\ref{eq:stableinterp}) is absolutely and uniformly convergent.  An
example of an interpolation kernel $\phi_\lambda$ mentioned in
Fact~\ref{fact:stableinterp} is the kernel $h$ in
Definition~\ref{def:zakaiclass} with $W = \pi$ and $\delta =
\pi(\lambda-1)$. In fact, there exist interpolation kernels
$\phi_{\lambda}$ that decay faster than $c_lt^{-l}$ for all $l \in
\nN$ and some constants $\{c_l\}_{l\in\nN}$. Technically, $\mathcal{T}
= \{t_l := l/\lambda\}_{l\in\zZ}$ is said to form a set of stable
sampling points for fields in $BL([-\pi,\pi])$
\cite{higginss1996}. This is because bounded perturbations in the
sample values at these locations never lead to unbounded
reconstruction errors at any point unlike in (\ref{eq:classicinterp})
where the sampling interval is exactly equal to a Nyquist-interval.

\begin{figure}[!htb]
\begin{center}
\scalebox{1.0}{\input{./Figures/classicNQ.pstex_t}}
\end{center}
\caption{\label{fig:classicNQ} \sl \small {\bf Classical
Nyquist-sampling with $3$-bit ADCs placed at Nyquist-locations
$\{l/\lambda\}_{l\in\zZ}$.} The entire budget of $k=3$ bits is
exhausted at a single sample point in any Nyquist-interval. The
spatially maximum interpolated reconstruction error decays
exponentially in the bits per Nyquist-interval $R = k\lambda$ with
exponent $(1/\lambda)$.}
\end{figure}
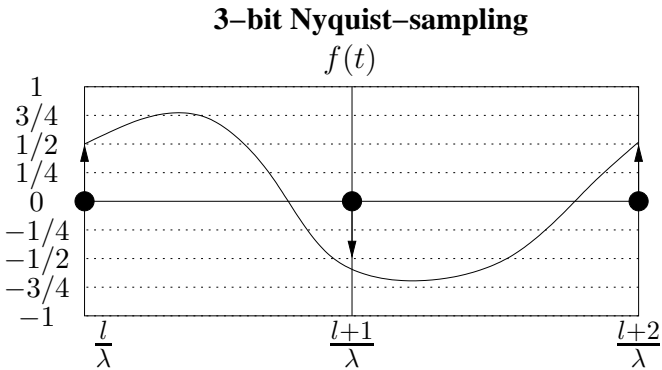

A ``near-Nyquist'' stable sampling rate $\lambda > 1$ but close to one
will be held fixed for the rest of this paper and the term
oversampling will be used to refer to uniform sampling at a spatial
sampling rate which is strictly greater than $\lambda$.  The term
``Nyquist-interval'' shall also be used, loosely, to refer to any
stable sampling interval of the form $[l/\lambda, (l+1)/\lambda), l\in
\zZ$.  The structure of Nyquist-sampling (for sampling rate $\lambda$)
is depicted in Figure~\ref{fig:classicNQ}.
To study the effect of finite precision ADCs on the reconstruction of
unit amplitude limited fields in $BL([-\pi,\pi])$, let $Q_k$ denote
the $k$-bit uniform scalar quantization operation on $[-1,1]$
\cite{GrayNIT98} so that $|z-Q_k(z)| \leq 2^{-k}$ for all
$z\in[-1,1]$.  The bitrate in bits per spatial Nyquist-interval for
each field snapshot used to quantize the field is given by $R =
k\lambda$.  Let the bitrate-$R$ quantized Nyquist-reconstruction of
unit amplitude limited $f \in BL([-\pi,\pi])$ be defined as:
\begin{eqnarray*}
\widehat{f}^{\mbox{\tiny NQ}}_R(t) &:=& \lim_{L\rightarrow\infty}
\sum_{l=-L}^{L}Q_k(f(l/\lambda))\phi\left(t - \frac{l}{\lambda}\right).
\end{eqnarray*}
Using (\ref{eq:stableinterp}) in Fact~\ref{fact:stableinterp} and the
triangle-inequality, it immediately follows that the spatially maximum
reconstruction error can be bounded as
\begin{eqnarray}
D(R) := ||f - {\widehat{f}}^{\mbox{\tiny NQ}}_R||_\infty &\leq&
C\cdot 2^{-\frac{1}{\lambda}R}.
\label{eq:NyqdetRDprofile}
\end{eqnarray}
Thus $D(R) = O(2^{-R/\lambda})$ as $R\rightarrow \infty$ and $R(D) =
O(\log_2(1/D))$ as $D\rightarrow 0$.  This shows that maximum error
can be made to decay to zero at least exponentially fast, with an
exponent $\lambda^{-1}$, as the bits per spatial Nyquist-interval goes
to infinity. The constant $C$ depends on the choice of interpolation
kernels.  We now argue, informally, that the rate of decay of the
maximum error cannot be much better than exponential in the
bitrate. Every sequence $\{z_l\}_{l\in\zZ}$ of values for the samples
at the Nyquist-locations, essentially determines an amplitude limited
$f$ in $BL([-\pi\lambda,\pi\lambda])$ via (\ref{eq:stableinterp}),
with $f(l/\lambda) \approx z_l$ for all $l\in\zZ$. Let
$\{z_l\}_{l\in\zZ}$ be a sequence whose quantization error magnitudes
$|z_l-Q_k(z_l)|$, $l\in\zZ$, are close to $2^{-k}$ and whose signs are
such that $\lim_{L\rightarrow \infty}\sum_{l=-L}^L(z_l -
Q_k(z_l))\phi_\lambda(t -l/\lambda)$ is close to $C2^{-k}$.  For such
an $f$ we have $D(R) \approx C\cdot 2^{-\frac{1}{\lambda}R}$. Thus
informally,
\begin{eqnarray*}
\sup_{f} ||f - {\widehat{f}}^{\mbox{\tiny NQ}}_R||_\infty &=&
O\left(2^{-\frac{1}{\lambda}R}\right).
\end{eqnarray*}
That is, for Nyquist-sampling with uniform scalar quantization and a
given choice of interpolation kernels, the maximum reconstruction
error for the worst amplitude limited field in $BL([-\pi,\pi])$ decays
exponentially in the bitrate. { A more precise statement
  of this result was formally established in
  \cite[Sec.~II.B]{DaubechiesDGV-AD:IT06} by computing the Kolmogorov
  $\epsilon$-entropy for the class of bounded square-integrable
  bandlimited functions.}

\subsection{Stochastic case}  \label{sec:NQstochext}

Let $X$ be as in Section~\ref{sec:stochfldmdl} (also see normalization
assumptions in the last paragraph of Section~\ref{sec:keytech}).  Then
in view of Fact~\ref{fact:stableinterp}, Lemma~\ref{lemma:Zakaiextn}
with $t_l = l\lambda^{-1}$ and $\psi_l(t) = \phi_\lambda(t)$,
$(l,t)\in \zZ\times \rR$, and Fact~\ref{fact:samplepaths},
almost-\-surely for all sample paths of $X$, we have $\forall t \in
\rR$,
\begin{eqnarray}
{
X(t) = 
\lim_{L\longrightarrow\infty} X_L(t), \ 
X_L(t) = \sum_{l=-L}^{L}X(l/\lambda)\phi_{\lambda} \left(
t-\frac{l}{\lambda}\right),
}
\label{eq:samprep}
\end{eqnarray}
where the above series converges absolutely for each $t$ and uniformly
on all compact subsets of $\rR$. In fact, for each $t \in \rR$, the
sequence of partial sums {$X_L$ in (\ref{eq:samprep})},
which are random variables, converges to $X(t)$ also in the expected
$p$-th power sense, $p\in (0,\infty)$, due to the following fact.

\begin{fact}({\em From almost-\-sure convergence to expected $p$-th power convergence}) \label{fact:as2ms}
{ If as $L \rightarrow \infty$, almost-\-surely $|X_L - X| \rightarrow
0$, and $|X_L| \leq A < \infty$ for all $L$ (then automatically $|X|
\leq A$ almost-\-surely), then by the { bounded
convergence theorem} for random variables~\cite[p.~17]{durrett},
$\lim_{L\rightarrow \infty}\eE|X_L - X|^p \rightarrow 0$ for all $p
\in (0,\infty)$.}
\end{fact}

For each $t$, let the bitrate-$R$ quantized Nyquist-reconstruction be
defined as:
\begin{eqnarray*}
\widehat{X}^{\mbox{\tiny NQ}}_R(t) &:=&
\lim_{L\rightarrow\infty}\sum_{l = -L}^{L}
Q_k\left(X\left(l/\lambda\right)\right)\phi\left(t -
\frac{l}{\lambda}\right),
\end{eqnarray*}
where for the same reasons as for (\ref{eq:samprep}), almost-\-surely
and in the expected $p$-th power sense, $p \in (0,\infty)$, the above
series converges absolutely for each $t$ and uniformly on all compact
subsets of $\rR$.  Thus as in (\ref{eq:NyqdetRDprofile}) we have,
almost-\-surely,
\begin{eqnarray}
D(R) = ||X - \widehat{X}^{\mbox{\tiny NQ}}_{R}||_\infty &\leq&
  C 2^{-\frac{1}{\lambda}R}.
\label{eq:NyqstocRDprofile}
\end{eqnarray}
This also implies that
\begin{eqnarray*}
||(\eE|X - \widehat{X}^{\mbox{\tiny NQ}}_{R}|^p)^{1/p}||_\infty &\leq&
  C 2^{-\frac{1}{\lambda}R},
\end{eqnarray*}
for all $p \in (0,\infty)$. As discussed in
Section~\ref{sec:distcrit}, the bound $C 2^{-\frac{1}{\lambda}R}$ in
(\ref{eq:NyqdetRDprofile}) and (\ref{eq:NyqstocRDprofile}) also
applies to the locally-averaged reconstruction errors. {
Since $R \propto k$ in the Nyquist-sampling framework, the
reconstruction quality can be improved only by using higher precision
ADCs.}

\Section{One-bit dithered oversampling (improving quality
through sensor density) \label{sec:ditherintro}}

In the Nyquist-sampling framework of the last section, samples are
collected at regular Nyquist-intervals and the ``bit-budget'' for each
interval is exhausted at a single sampling location. In this
framework, the reconstruction accuracy can be improved only by
improving the precision of the ADCs of the sensors. High-precision
ADCs are expensive and have higher power requirements. Future sensor
networks are envisioned to be made of cheap, low-power, low-precision
devices deployed in large numbers over a given geographical area of
interest. The question which arises naturally in this context is
whether it is possible { to} realize high-resolution field
reconstruction using a dense network of low-precision sensors. Can the
fixed sensor-precision be adequately compensated through high
sensor-density?  If so, how does the reconstruction quality scale with
the sensor density? In particular, can the maximum reconstruction
error over space be made to decay exponentially with the bit-budget
per Nyquist-interval as in the Nyquist-sampling framework? The answers
to these questions will help characterize the ``information density''
in bits per meter associated with sampling a bandlimited field
independent of sensor-precision.  The focus of this section is on
high-resolution field reconstruction using one-bit-precision sensors
in the deterministic and stochastic cases. We first summarize the
results for the deterministic case due to Cvetkovi\'{c}, and
Daubechies from \cite{zorands2000} and then develop the stochastic
counterpart.  In the next section we study the general case of field
reconstruction with $b$-bit-precision sensors in the deterministic and
stochastic cases and expose an underlying ``conservation of bits''
principle.

There is a vast body of literature available on classical one-bit
oversampled analog-to-digital conversion. This includes (i)
oversampled sigma-delta modulation used in audio consumer equipment
for many years
\cite{Gray-O:TransCom87,ThaoV-RO:TransSP94,ThaoV-DA:TransSP94,ZoranV-ER:IT98},
and (ii) oversampled randomized dithered averaging
\cite{MasryC-CE:IT81,Masry-TR:IT81,Marvasti-AU:NonUniform87,Zayed93,Marvasti-NS:Plenum01}
which has connections with stochastic resonance theory in physics,
nonlinear dynamical systems, neural sensory information processing
\cite{stochres1,stochres2,stochres3}, and has been used for real-time
synthetic-aperture-radar imaging
\cite{PascazioS-SA:ElecCommEnggJrnl98,FranceschettiTR-OB:IGARSS99}
(also see and references therein). One limitation of methods like
sigma-delta modulation is that they are not amenable to distributed
implementation because sampling is done in a sequential manner where
the result of quantization at the previous sampling location needs to
be available at the next location for quantization. A second
limitation is that in all these methods, the bitrate is proportional
to the oversampling rate. As a result, they have a {\em polynomial}
distortion-rate decay characteristic, example, $D(R) = O(R^{-(m+1)})$
for $m$-stage oversampled sigma-delta conversion
\cite{Gray-O:TransCom87,ThaoV-RO:TransSP94,ThaoV-DA:TransSP94} where
$D(R)$ is the maximum reconstruction error and $R$ is the bit-budget
per Nyquist-interval. Hence, there is a fundamental {\em exponential
performance gap} between classical methods and Nyquist-sampling
(compare with (\ref{eq:NyqdetRDprofile})).

{A key idea for achieving an exponential distortion-rate
decay characteristic, as in (\ref{eq:NyqdetRDprofile}), using
fixed-precision ADCs, is dithered oversampling.  This idea was first
proposed by Cvetkovi{\'c} and Daubechies in \cite{zorands2000} for the
deterministic case using single-bit ADCs.  The details of this idea
and our extensions to the stochastic case are described in the
following subsections. Dithered oversampling with multi-bit ADCs is
discussed in the next section.}

\subsection{Deterministic case}\label{sec:onebitdithdet}

Let $f$ be any unit amplitude limited field in $BL([-\pi,\pi])$.  The
key component of dither-based sampling schemes is the dither field
$d(t)$ that has the following properties:
\begin{enumerate}
\item $\forall l \in \zZ,\quad 1 < \gamma := |d(l/\lambda)| < \infty$;
\item $\forall l\in\zZ, \quad \mbox{sign}[d(l/\lambda)] = -\mbox
  {sign} [d((l+1)/\lambda)]$;
\item $d(t)$ is { continuous everywhere and differentiable
everywhere except possibly in
$\left\{\left({l}/{\lambda}\right)\right\}_{l\in\zZ}$ where it is left
and right differentiable. All differentials are uniformly bounded in
magnitude by $\Delta \in (0, \infty)$.}
\end{enumerate}
For example, $d(t) = \gamma \cos(\lambda \pi t)$, $t\in\rR$, with
$|\gamma| > 1$ is a valid dither field.  { Sensors ``add''
the value of a pre-designed smooth dither field $d(t),t\in\rR$, at
their respective locations to the value of their observations $f(t)$
and note only the {\em sign} of the sum. In terms of hardware, this
can be implemented using threshold comparators based on operational
amplifiers.}  
Since $f$ is continuous, the dithered field $[f + d]$ is
continuous. The first two properties of $d$ and the unit amplitude
limit on $f$ guarantee that $[f+d](l/\lambda)$ and
$[f+d]((l+1)/\lambda)$ will have opposite signs. By the intermediate
value theorem for continuous functions \cite{rudinp1976}, it follows
that $[f+d]$ will have a zero-crossing (at least one but possibly
more) in every open Nyquist-interval $(l/\lambda,(l+1)/\lambda)$.

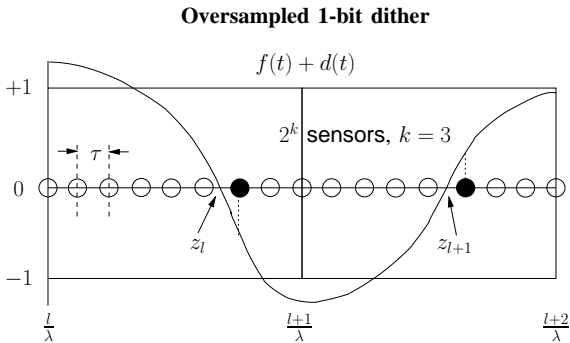
\begin{figure}[!htb]
\begin{center}
\scalebox{0.5}{\input{Figures/1bitdither.pstex_t}}
\end{center}
\caption{\label{fig:1bitdither} \sl \small {\bf One-bit
dithered sampling with $8$ one-bit sensors separated by $\tau =
1/(8\lambda)$.} 
The locations of 
the first zero-crossing (from the left) in each Nyquist-interval are
denoted by $z_l$ in the figure.
The spatially maximum interpolated reconstruction error decays
exponentially in $k$ with exponent $(1/\lambda)$ as in the Nyquist
sampling framework.}
\end{figure}

Let $N = 2^k$ ($k \geq 1$), one-bit ADCs be placed uniformly in every
Nyquist-interval to record the sign of the dithered field $[f+d]$,
that is, sensors are placed at the locations $\tau_k\zZ$ where $\tau_k
:= (1/\lambda)2^{-k}$ is the uniform oversampling
period. { Typically $\tau \ll 1/\lambda$}.  To avoid
clutter, we shall henceforth drop the subscript $k$ in $\tau_k$. Let
$m_l \in \{0,\ldots,(N-1)\}$ be the {\em smallest} index for which
$[f+d]((l/\lambda)+m_l\tau)$ and $[f+d]((l/\lambda)+(m_l+1)\tau)$ have
opposite signs in $(l/\lambda, (l+1)/\lambda)$. It follows from the
intermediate value theorem that for each $l\in \zZ$, $f(z_l) + d(z_l)
= 0$ at some point $z_l \in
((l/\lambda)+m_l\tau,(l/\lambda)+(m_l+1)\tau)$.  To avoid ambiguity,
let $z_l$ be the location of the {\em first} (from left to right)
zero-crossing of $[f+d]$ in the interval $[l/\lambda,(l+1)/\lambda]$.
The structure of one-bit dithered sampling is depicted in
Figure~\ref{fig:1bitdither}.
Uniformly amplitude limited and bandlimited fields have uniformly
amplitude limited derivatives. Specifically, according to Bernstein's
inequality (\ref{eq:Bernstein}) in Section~\ref{sec:fieldmodel}, for
unit amplitude limited fields in $BL([-\pi,\pi])$, $\forall t\in\rR$,
$|f'(t)| \leq \pi$.  The third condition on the dither field and
(\ref{eq:Bernstein}) together ensure that the value of the derivative
of $\left[f + d\right]$ is uniformly bounded by $\left(\pi +
\Delta\right)$.  Applying Lagrange's mean value theorem
\cite{rudinp1976} to the mid-point of the ``uncertainty zone'' $t_l :=
((l/\lambda) + (m_l+(1/2)) \tau )$ and $z_l$, we get
\begin{eqnarray*}
\left|[f+d](z_l) - \left[f+d\right]\left(t_l\right) \right| \leq (\pi
+ \Delta) \left|z_l - t_l\right|,\quad\mbox{that is,} \\
\left|f\left(t_l\right) -
 \left(-d\left(t_l\right)\right)\right| \leq \left(\frac{\pi + \Delta}{2}\right)\tau = \left(\frac{\pi +
\Delta}{2\lambda}\right) 2^{-k}.
\end{eqnarray*}
Thus, uniform oversampling of the dithered field using one-bit ADCs
gives samples of $f$ having linear precision in $\tau$ (exponential in
$k$) at the {\em nonuniformly} spaced locations $\mathcal{T} =
\{t_l\}_{l\in\zZ}$ which depend on both the field and the dither.
{ Loosely speaking, the dither field provides a way of
translating spatial resolution/uncertainty of the zero-crossing
locations (but amplitude certainty, equal to zero) to amplitude
resolution/uncertainty of the field at these locations (but spatial
certainty, equal to the mid-point of the uncertainty zone).}

{ Unlike Nyquist-sampling where the bit-budget for each
Nyquist-interval is exhausted at a single sampling point, the
dither-based approach ``spreads'' the bit-budget over many single-bit
ADCs in a Nyquist-interval.}  There are $N = 2^k$ one-bit sensors
uniformly distributed over an interval of length $(1/\lambda)$. It
takes $k$ bits or a bitrate of $R = k\lambda = O(\log N)$ bits per
Nyquist-interval per field snapshot to index the location of the first
zero-crossing.
{ Deferring a full discussion of distributed processing
and communication issues and costs to
Section~\ref{sec:sensornetworks}, as a conceptual aid one may imagine
that when a sensor observes a sign change relative to its neighbors,
it can wirelessly broadcast its identity to the CPU using $\log N$
bits with all other sensors remaining silent.
}
For each $l\in\zZ$, the sample errors decay with rate $R$ as follows
\begin{eqnarray}
\left|f\left(t_l\right) - \left(-d\left(t_l\right)\right)\right| &\leq
& \left(\frac{\pi+\Delta}{2\lambda}\right)
2^{-\frac{1}{\lambda}R}. \label{eq:sampRD1bitdith}
\end{eqnarray}

{ This per-sample amplitude resolution, which is
proportional to inter-sensor separation, can be converted to global
field amplitude resolution of the same order using the following
result on field interpolation from nonuniformly spaced samples
(adapted from \cite[Theorem~3.2]{zorands2000}) due to Cvetkovi{\'c}
and Daubechies.}

\begin{fact} ({\em Stable interpolation of bandlimited fields from samples
  at nonuniformly spaced locations \cite[Theorem~3.2]{zorands2000}})
\label{fact:zoransmainthm} {Let $\lambda > 1$ and $\mathcal{T} =
  \{t_l\}_{l\in\zZ}$ be a set of sampling locations such that
$\underline{\kappa}:=\inf_{j,l\in\zZ,j\neq l} \left|t_j - t_l\right| >
0$ and $\bar{\kappa} :=\sup_{l\in\zZ} |t_l - (l/\lambda)| < \infty$.
Then there exist absolutely-\-integrable interpolation kernels
$\psi_l$ belonging to $BL([-\pi\lambda,\pi\lambda])$ such that $C' :=
\sup_{t\in\rR}\left[\sum_{l\in\zZ}|\psi_l(t-t_l)|\right] < \infty$ and
for all fields $f \in BL({[-\pi\lambda + \delta,\pi\lambda
- \delta]})$ with $0 < \delta < {\pi\lambda}$,
\begin{eqnarray*}
f(t) = \lim_{L\longrightarrow\infty}\sum_{l=-L}^{L} f(t_l) \psi_l(t-t_l),
\end{eqnarray*}
where the above series converges absolutely for each $t$ (and in
$\mathcal{L}^2$) and uniformly on all compact subsets of $\rR$. The
interpolation kernels $\psi_l$ depend on the sampling set
$\mathcal{T}$ but there exist interpolation kernels for which $C'$
depends only on $\lambda$, $\underline{\kappa}$, and $\bar{\kappa}$
and not on the specific sampling set $\mathcal{T}$.}
\end{fact}

In essence, the field can be reconstructed in a stable manner as in
the Nyquist-sampling framework as long as the sampling locations do
not get too close to each other (not closer than $\underline{\kappa} >
0$) and do not run away too far from the Nyquist-points (not farther
than $0 < \bar{\kappa} < \infty$). There exist interpolation kernels
$\psi_l$ that decay faster than $(c_m'/t^m)$ for all $m$ {
where the constants $\{c_m'\}_{m\in\nN}$ depend only on $\lambda$,
$\underline{\kappa}$, and $\bar{\kappa}$ and not on
$\mathcal{T}$~\cite{zorands2000}.} 
As discussed in Section~\ref{sec:DataTran}, these properties prove
useful in the practical context of reconstructing the bandlimited
field to a desired {\em nonzero} distortion from samples collected in
only a {\em finite} spatial region of interest.

In Remark~\ref{rem:techpoint1} below, it is shown that $\mathcal{T} =
\{t_l := ((l/\lambda) + (m_l+(1/2)) \tau )\}_{l\in\zZ}$ satisfies the
conditions of Fact~\ref{fact:zoransmainthm} { with
$\underline{\kappa}$, $\bar{\kappa}$, and $C'$ which do not depend on
$\tau$.}  Hence, from (\ref{eq:sampRD1bitdith}),
Fact~\ref{fact:zoransmainthm}, and the triangle inequality, it follows
that there exist interpolation kernels $\{\psi_l\}_{l\in\zZ}$ (which
depend on $\mathcal{T}$ in general) for which the spatially maximum
interpolated reconstruction error is bounded as follows (also see
\cite[Corollary~3.3]{zorands2000})
\begin{eqnarray}
\label{eq:one-bitdetRDprofile}
D(R) := ||f-\widehat{f}^{\mbox{\tiny one-bit}}_R||_{\infty} &\leq&
 C'\left(\frac{\pi + \Delta}{2\lambda}\right)
 2^{-\frac{1}{\lambda}R}, \\  \mbox{where} \quad \widehat{f}^{\mbox{\tiny
 one-bit}}_R(t) &:=&
 \sum_{l\in\zZ}\left(-d\left(t_l\right)\right)\psi_l\left(t-t_l\right). \nonumber
\end{eqnarray}
The reconstruction error decays exponentially with the bitrate, with
the same exponent as in the Nyquist-sampling scheme
(\ref{eq:NyqdetRDprofile}).  The reconstruction accuracy can be
improved by reducing $\tau$, that is, by packing more sensors inside
each Nyquist-interval. However, unlike the Nyquist-sampling scheme,
there is no need to use higher precision ADCs.

\begin{remark}{\rm \label{rem:techpoint1}
We are interested in the behavior of the reconstruction error as the
number of sensors $N = 2^k = 2^{R/\lambda}$ increases. Since $C'$
depends on $\underline{\kappa} =\inf_{j\neq l} |t_j -t_l|$ and
$\bar{\kappa} :=\sup_{l\in\zZ} |t_l - (l/\lambda)|$ (but the
interpolation kernels $\psi_l$s can be designed so that $C'$ does not
depend on the specific nonuniform sampling set), to truly assert that
the decay of the reconstruction error is exponential in $R$, we must
ensure that $\underline{\kappa}$ and $\bar{\kappa}$ do not depend on
$\tau = (1/\lambda)2^{-k}$. Clearly, $\bar{\kappa} < 1/\lambda$. What
about $\underline{\kappa}$? A lower bound on $\underline{\kappa}$ is
$\tau$ since the $\{t_l\}_{l\in\zZ}$ cannot get closer than $\tau/2$
to the endpoints of the Nyquist-interval by design. This lower bound
depends on $\tau$. However, since $|d(t) + f(t)| \geq |d(t)| - |f(t)|
\geq |d(t)| - 1$, a zero-crossing cannot occur unless the amplitude of
$d$ falls below unity. Since $|d(l/\lambda)| = \gamma > 1$, the
closest that a zero-crossing can get to a Nyquist-point is
$(\gamma-1)/\Delta$ where $\Delta$ is the maximum slope of $d$. Since
the $t_l$s are midpoints of inter-sensor intervals,
$\underline{\kappa} \geq (2(\gamma-1)/\Delta) - \tau$. Together with
the earlier lower bound $\underline{\kappa} \geq \tau$, this means
that $\underline{\kappa} \geq (\gamma-1)/\Delta$ which is independent
of $\tau$. For aiding comparison with $b$-bit dithered oversampling of
the next section, we take $C'$ to correspond to the conservative
values $\bar{\kappa}^{\mbox{\tiny one-bit}} = 1/\lambda$ and
$\underline{\kappa}^{\mbox{\tiny one-bit}} =
\min(1/(2\lambda),(\gamma-1)/\Delta)$.}
\end{remark}

\begin{remark}{\rm
From (\ref{eq:one-bitdetRDprofile}), we would like the maximum slope
$\Delta$ of the dither field to be as small as possible to get a
tighter upper bound on the error. The average slope of the dither
field in any Nyquist-interval is $\left(2\gamma\lambda\right)$. Since
$d$ is smooth on every open Nyquist-interval, by the mean-value
theorem, there is always some point in the open interval where the
slope of $d$ matches the average slope. Hence, $\Delta$ can be no
smaller than $ \left(2\gamma\lambda\right)$. We can make $\Delta =
\left(2\gamma\lambda\right)$ by using a piece-wise linear dither
field.}
\end{remark}

\begin{remark}{ \label{rem:techpoint2} \rm
It is in fact possible to design a dither field $d$ for which $[f+d]$
will have {\em exactly one} zero-crossing in each Nyquist-interval:
consider for example the continuous periodic triangular dither field
which is piecewise linear in each Nyquist-interval and whose slope
$\Delta$ is strictly greater than $\pi$ in magnitude. For such a $d$,
the sum field $[f+d]$ is strictly monotonic in each Nyquist-interval
because the slope of the sum field never changes sign in any Nyquist
interval. { For expositional simplicity and to aid the
discussion of communication issues and costs in
Section~\ref{sec:sensornetworks}}, for the remainder of this paper, we
will assume that the dither field satisfies this property.}
\end{remark}

\subsection{Stochastic case} \label{sec:onebitstochext}

Let $X$ be as in Section~\ref{sec:stochfldmdl} (also see normalization
assumptions in the last paragraph of Section~\ref{sec:keytech}).  It
should be noted that $X$ is a random process and that its sample paths
are not bandlimited in the conventional sense. There may even be
discontinuous sample paths (having probability zero) for which $[X+d]$
may not have an actual zero-crossing. However, since $|X| \leq 1$ and
$\gamma > 1$, $\left[X+d\right](l/\lambda)$ and
$\left[X+d\right]((l+1)/\lambda)$ will always have opposite
signs. Hence, in each Nyquist-interval, there always exists a first
sensor which will detect a {\em sign-change}. As in the deterministic
case, let $M_l$ be the index (a discrete random variable) of the
sensor located just left of the first sign-change and $T_l :=
[(l/\lambda) + (M_l+(1/2)) \tau ]$. Note that the set $\mathcal{T} =
\{T_l\}_{l\in\zZ}$ is a countable set of random ``zero-crossings''
(more accurately, first sign-change locations) which is closely
coupled to the underlying spatial stochastic process $X$.  However,
each $T_l$ can take values only in the deterministic discrete set
$\{(l/\lambda) + (m+0.5)\tau:~m = 0,\ldots,(N-1)\}$.

From Fact~\ref{fact:samplepaths} and the Remark~\ref{rem:techpoint1},
almost-\-surely the random sampling set $\mathcal{T}$ satisfies the
conditions of Fact~\ref{fact:zoransmainthm}. Then from
Fact~\ref{fact:zoransmainthm} and Lemma~\ref{lemma:Zakaiextn} with $C
= C'$, almost-\-surely for all sample paths of $X$, $\forall t \in
\rR$,
\begin{eqnarray}
X(t) &=& \lim_{L\longrightarrow\infty}
\sum_{l=-L}^{L}X(T_l)\psi_l\left(t-T_l\right),
\label{eq:samprep2}
\end{eqnarray}
where the above series converges absolutely for each $t$ and uniformly
on all compact subsets of $\rR$. In fact, due to Fact~\ref{fact:as2ms},
for each $t \in \rR$, the sequence of partial sums to the right of
(\ref{eq:samprep2}), which are random variables, converges to $X(t)$
also in the expected $p$-th power sense, $p\in (0,\infty)$.

From Fact~\ref{fact:samplepaths} and Proposition~\ref{prop:bddslope},
almost-\-surely, all sample paths of $X$ are differentiable
everywhere, and the derivative $X'$ has a normalized amplitude which
is uniformly bounded by $\pi$. Hence, by (\ref{eq:sampRD1bitdith}),
almost-\-surely for all paths of $X$,
\begin{eqnarray}
\label{eq:stocsampRD1bitdith}
\left|X(T_l)-\left(-d\left(T_l\right)\right)\right| &\leq&
 \left(\frac{\pi + \Delta}{2\lambda}\right)
 2^{-\frac{1}{\lambda}R},
\end{eqnarray}
$\forall l\in\zZ$.  For each $t\in\rR$, let the bitrate-$R$, one-bit
quantized reconstruction of $X$ be defined as:
\begin{eqnarray}
\widehat{X}^{\mbox{\tiny one-bit}}_R(t) &:=&
\sum_{l\in\zZ}\left(-d\left(T_l\right)\right)\psi_l\left(t-T_l\right),
\label{eq:defstocdithrec}
\end{eqnarray}
where for the same reasons as for (\ref{eq:samprep2}), almost-\-surely
and in the expected $p$-th power sense, $p \in (0,\infty)$, the above
series converges absolutely for each $t$ and uniformly on all compact
subsets of $\rR$.  Thus as in (\ref{eq:one-bitdetRDprofile}),
almost-\-surely,
\begin{eqnarray}
D(R) = ||X - \widehat{X}^{\mbox{\tiny one-bit}}_R||_\infty 
&\leq& C'' 2^{-\frac{1}{\lambda}R},
\label{eq:one-bitstocRDprofile}
\end{eqnarray}
with $C'' := C'(\pi + \Delta)/(2\lambda)$. This also implies that
\begin{eqnarray*}
||(\eE|X - \widehat{X}^{\mbox{\tiny one-bit}}_{R}|^p)^{1/p}||_\infty&\leq& C'' 2^{-\frac{1}{\lambda}R},
\end{eqnarray*}
for all $p\in(0,\infty)$.  Once again, as discussed in
Section~\ref{sec:distcrit}, the bound $C'' 2^{-\frac{1}{\lambda}R}$ in
(\ref{eq:one-bitdetRDprofile}) and (\ref{eq:one-bitstocRDprofile})
also applies to the locally-averaged reconstruction errors. It should
be noted that $C'$ depends only on $\lambda, \underline{\kappa}$, and
$\bar{\kappa}$, which are fixed, and not on the specific realization
of the set of random sampling locations $\mathcal{T}$ which satisfy
the assumptions of Fact~\ref{fact:zoransmainthm}. However, in general
the set of interpolation kernels $\{\psi_l\}_{l\in\zZ}$ depends on the
{\em random} sampling set $\mathcal{T}$.

{\noindent{\em Discussion:}} In the stochastic one-bit
dithered oversampling scenario of this and the following section, the
sampling locations $\mathcal{T}$ and the associated interpolation
kernels are tightly coupled to the underlying process $X$ because they
arise as the locations of level-crossings of the process $X$. Most
results in the nonuniform sampling literature (see
\cite{Marvasti-AU:NonUniform87,Zayed93,Marvasti-NS:Plenum01} and
references therein) address the reconstruction of (i) deterministic
bandlimited fields from nonuniform deterministic {sampling
locations}, (ii) bandlimited stochastic processes from nonuniform
deterministic {sampling locations}, (iii) deterministic
bandlimited fields from random {sampling locations}, and
(iv) bandlimited stochastic processes from {\em independent} random
{sampling locations}. Scenarios (iii) and (iv) may arise,
for example, due to random sensor-deployment independent of the field
being sensed. However, the results here are for the reconstruction of
WSS bandlimited stochastic processes from random {sampling
locations} which are {\em strongly dependent} on the process being
sampled.

Summarizing (\ref{eq:NyqdetRDprofile}), (\ref{eq:NyqstocRDprofile}),
(\ref{eq:one-bitdetRDprofile}), and (\ref{eq:one-bitstocRDprofile}),
it is clear that for a fixed $\lambda, \gamma > 1$, the
$\mathcal{L}^\infty$ norm of the point-wise error in the
Nyquist-reconstruction or one-bit dithered reconstruction decays at
least exponentially fast in the bitrate (bits per spatial
Nyquist-interval for each field snapshot) $R$ with exponent
$(1/\lambda)$. Similar rate-error characteristics hold for the
deterministic and expected local-average distortion criteria of
Section~\ref{sec:models}. Hence, if $D$ denotes the maximum point
error, the ``intrinsic information density'' (ignoring data-transport
issues) $R(D)$ in bits per meter grows no faster than (with the
almost-\-sure/expected qualifiers in the stochastic case) $R(D) =
O\left(\log(1/D)\right)$ as $D$ decreases to $0$. For one-bit dithered
sampling, the number of sensors per Nyquist-interval $N$, grows no
faster than $O(1/D)$, or equivalently, $D(N) = O(1/N)$. The total
information in bits per Nyquist-interval grows no faster than $O(\log
N)$ and the bits per sensor decreases at least as fast as $O((1/N)\log
N)$.

\Section{Tradeoffs in amplitude and spatial resolution: a
``bit-conservation'' principle}\label{sec:bcp} 

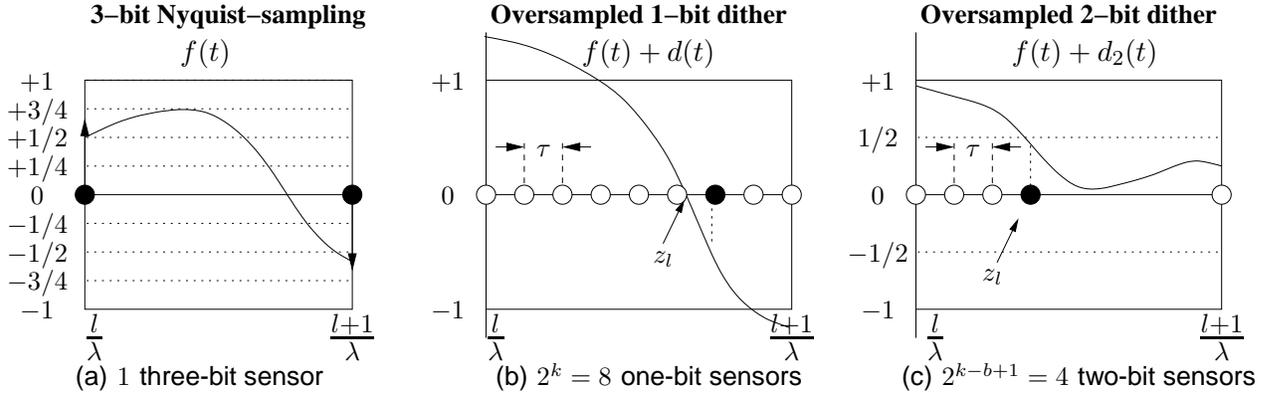
\begin{figure*}[!htb]
\begin{center}
\scalebox{1.0}{\input{./Figures/bcp.pstex_t}}
\end{center}
\caption{\label{fig:bbitdithereg} \sl \small {\bf Amplitude resolution
versus oversampling rate tradeoffs in conventional and dither-based
sampling frameworks.} (a) Conventional (Nyquist) sampling using
$3$-bit ADCs placed at Nyquist-locations
$\{l/\lambda\}_{l\in\zZ}$. The entire budget of $3$ bits is exhausted
at a single sample point in any Nyquist-interval. (b) Dither-based
sampling scheme using $8$, one-bit ADCs distributed uniformly over a
Nyquist-interval to locate the zero-crossing $z_l$ of the dithered
field. (c) A flexible tradeoff between these extremes. Four, $2$-bit
ADCs distributed uniformly over \underline{half} the Nyquist-interval
detect a level-crossing $z_l$ at level $+(1/2)$. All three schemes
have similar exponential error accuracy in the bitrate.}
\end{figure*}

In Nyquist-sampling, fields are sampled at ``low'',
near-Nyquist-rates.  The entire bit-budget of $k$ bits per
Nyquist-interval is spent in recording the field amplitude at a single
high-precision ($k$-bit) sensor. On the other hand, the one-bit
dither-based sampling scheme spends all available bits in recording
the location of a zero-crossing (a spatial event) by using many ($N =
2^k$) poor precision (one-bit) sensors in each Nyquist-interval. These
sampling schemes represent two extreme scenarios.  This section
explains how $k$-bit Nyquist-sampling accuracy can be achieved using
$b$-bit ADCs and an appropriate dither-based oversampling scheme for
any $1 < b < k$. This leads to a bit-conservation principle -- a
trade-off between the oversampling factor and ADC-precision for
``similar'' asymptotic reconstruction accuracy. To the best of our
knowledge, this tradeoff has not been discussed before in the sampling
literature. We discuss this in detail for deterministic fields. The
extension to the stochastic setting parallels the stochastic
extensions in Sections~\ref{sec:NQstochext} and
\ref{sec:onebitstochext}.

\subsection{Deterministic case} 

Nyquist-sampling uses only one $k$-bit ADC per Nyquist-interval of
length $1/\lambda$. The one-bit dithered sampling scheme uses $N =
2^k$, one-bit ADCs distributed uniformly over the same interval, that
is, the ADCs are placed at intervals of length $\tau = 1/(N
\lambda)$. For definiteness, assume that the sensors are placed at the
beginning of every $\tau$-length interval, that is, at locations
$\{m\tau\}_{m\in\zZ}$. Now consider the scenario where
{$N_b := 2^{k-b+1}$} $b$-bit ADCs ($1 < b < k$) are placed
{$\tau$ apart, in the leftmost section of the Nyquist
interval}
Notice that, one-bit ADCs only detect one level-crossing (the $0$
level), 2-bit ADCs can detect 3 distinct level-crossings, and in
general, $b$-bit ADCs can detect $(2^b - 1)$ distinct level-crossings
given by $\{0,\pm(1/2^{b-1}),\ldots,\pm(1-(1/2^{b-1})) \}$.

Appendix~\ref{ap:bbitdither} shows how to design a $b$-bit dither
field $d_b$, with maximum slope magnitude $\widetilde{\Delta}$ {\em
independent of} $b$, so that $[f + d_b]$ always crosses some
quantization level in every interval of the form $[A_l,B_l] :=
[l/\lambda, {(l + (N_b -1)/N)/\lambda}] \subset
[l/\lambda, (l+1)/\lambda]$.\footnote{In fact, it is possible to
construct $d_b(t)$ using any dither field $d(t) = d_1(t)$ of the
one-bit sampling scheme via Equation~(\ref{eq:bfrom1}) in
Appendix~\ref{ap:bbitdither}. Then, the maximum slope of the $b$-bit
dither field in any interval $(A_l,B_l)$ is no more than
$\widetilde{\Delta} = (4c/\gamma)\Delta$, $c > (1+\pi)/4$, where
$\Delta$ is the maximum slope of the one-bit dither in any interval
$(l/\lambda,(l+1)/\lambda)$.} Each interval $[A_l,B_l]$
{usually} covers only a small fraction of the length of a
Nyquist-interval and contains $N_b$ $b$-bit ADCs. The situation is
graphically illustrated in Figure~\ref{fig:bbitdithereg}.

Let $m_l \in \{0,\ldots,{(N_b-2)}\}$ be the smallest index
for which $[f+d_b](A_l + m_l \tau)$ and $[f+d_b](A_l + (m_l+1)\tau)$
are on opposite sides of some level $q_l \in
\{0,\pm(1/2^{b-1}),\ldots,\pm(1-(1/2^{b-1}))\}$ and let $z_l \in
(A_l+m_l\tau,A_l+(m_l+1)\tau)$ be the actual point of level-crossing,
that is, $f(z_l) + d_b(z_l) = q_l$.  From the mean value theorem
applied to $[f+d_b](t)$ for the end-points $t_l := A_l + (m_l +
(1/2))\tau$ and $z_l$ we obtain
\begin{eqnarray*}
\left|\left[f+d_b\right](z_l) -
\left[f+d_b\right]\left(t_l\right)\right| \leq \sup_{t\in
(A_l,B_l)}\left|\left[f'+d_b'\right](t)\right| \left|z_l - t_l\right|,
\end{eqnarray*}
which leads to,
\begin{eqnarray*}
\left|f\left(t_l\right) - \left(q_l
- d_b\left(t_l\right)\right)\right| &\leq &
\left(\frac{\pi + \widetilde{\Delta}}{2} \right) \tau.
\end{eqnarray*}
This shows that the accuracy of the nonuniform samples
$\{f\left(t_l\right)\}$ is linear in $\tau$, {\it independent} of the
precision of the ADCs just as in the one-bit dithered sampling scheme.

It requires only {$\lceil \log_2({N_b} -
  1)\rceil \leq (k-b+1)$} bits per Nyquist-interval to specify the
location of the sensor {\it just} following the location of the first
level-crossing of $[f+d_b](t)$.  It takes no more than $\log_2(2^b -
1) < b$ bits to index the level that was crossed first in each
Nyquist-interval.
Hence, the total number of bits required with this sampling method
is not more than $(k - b + 1) + b = (k+1)$ bits (or $R = \lambda
(k+1)$ bits per interval) which 
{ gives the same distortion-rate decay characteristics as
the} $k$-bit Nyquist-sampling and the one-bit dither-based sampling
schemes.
{ As in Section~\ref{sec:ditherintro}, we defer a full
discussion of communication issues and costs to
Section~\ref{sec:sensornetworks}. As a conceptual aid, however, one
may imagine that the sensor observing a level-crossing relative to its
neighbor wirelessly broadcasts its index and the index of the level
that has been crossed to the CPU using $(k+1)$ bits with all other
sensors remaining silent.}
The source of this additional bit can be explained as follows: In
one-bit dithered sampling, the ADCs do not need to explicitly specify
which level was crossed since there is only one level; only
zero-crossing locations need to be described. In the $k$-bit
Nyquist-sampling, the locations of the ADCs need not be explicitly
specified since they are known; however, the sensors need to specify
the quantization interval in which the field sample lies.

Since $\tau$ and $k$ are related via $\tau = 1/(\lambda 2^k)$,
therefore, $\tau = (2/\lambda)2^{-\frac{1}{\lambda}R}$, and the
maximum sample error is no more than
\begin{eqnarray}
|f(t_l) - \left(q_l-d_b(t_l)\right)| &\leq& \left(\frac{\pi +
 \widetilde{\Delta}}{\lambda}\right) 2^{-\frac{1}{\lambda}R}, 
\label{eq:precisionofsamples}
\end{eqnarray}
$\forall l\in \zZ$.  Following the analysis for the one-bit case, it
can be verified that $\{t_l\}_{l\in\zZ}$ satisfies the conditions of
Fact~\ref{fact:zoransmainthm}. Specifically, $\bar{\kappa} \leq
(B_l-A_l) = 1/(\lambda 2^{(b-1)}) \leq 1/\lambda =
\bar{\kappa}^{\mbox{\tiny one-bit}}$ and $\underline{\kappa} \geq
(A_{l+1}-B_l) \geq (1 - (1/2^{(b-1)})) (1/\lambda) \geq 1/(2\lambda)
\geq \underline{\kappa}^{\mbox{\tiny one-bit}}$ (see
Remark~\ref{rem:techpoint1}).  Hence, if the rate-$R$, $b$-bit
reconstruction is given by $\widehat{f}^{\mbox{\tiny $b$-bit}}_{R}(t)
:= \sum_{l\in\zZ}\left(q_{l} -
d_b\left(t_l\right)\right)\psi_l\left(t-t_l\right)$,
\begin{eqnarray}
||f-\widehat{f}^{\mbox{\tiny $b$-bit}}_{R}||_{\infty} \leq
\widetilde{C}'\left(\frac{\pi + \widetilde{\Delta}}{\lambda}\right)
2^{-\frac{1}{\lambda}R},
\label{eq:bbitdetRDprofile}
\end{eqnarray}
where the constant $\widetilde{C}'$ (tilde is used to distinguish the
constants of this section from those in Section~\ref{sec:ditherintro})
does not depend on $\tau$, $f$, and the individual $\{t_l\}_{l\in\zZ}$
but only on $\underline{\kappa}$, $\bar{\kappa}$, and $\lambda$. If we
use the conservative values $\bar{\kappa} = \bar{\kappa}^{\mbox{\tiny
one-bit}}$ and $\underline{\kappa} = \underline{\kappa}^{\mbox{\tiny
one-bit}}$ then $\widetilde{C}' = C'$.  Hence, independent of $b$, the
distortion-rate asymptotics are the same as in the Nyquist-sampling
and the one-bit dithered sampling set-ups, that is, the distortion is
exponentially decaying in rate $R$ with the same exponent. This
observation is summarized in the following principle. \\

\noindent{\bf ``Conservation of bits'' principle}: Let $k$ be the
bit-budget in terms of the total number of bits available per
Nyquist-interval. For each $1\leq b < k$ there exists a (dither-based)
sampling scheme with not more than $2^{k-b+1}$, $b$-bit sensors per
Nyquist-interval achieving a spatially maximum reconstruction error
$D$ of the order of $2^{-k}$. Alternatively, if there are $N_b$,
$b$-bit sensors in each Nyquist-interval, then $D = O(2^{-(b + \log_2
N_b)})$. The maximum error decays exponentially in the bit-budget.
The bit-budget can be apportioned in a flexible manner between
amplitude resolution in bits $b$ and the spatial resolution in bits
$\log_2 N_b$. It should be noted that $N_b$ is the sensor density in
terms of the number of sensors per Nyquist-interval.

\subsection{Stochastic case} \label{sec:bcpstochext}

The procedure for extending the results from the deterministic to the
stochastic $b$-bit setting is identical to the stochastic extension of

the one-bit deterministic dither scheme.  $X$ is sampled by adding
$d_b$ having the same properties as in the deterministic case. Let
$M_l$ be the index of the sensor just prior to the first level-change,
$T_l := [l/\lambda + (M_l + (1/2))\tau ]$ the mid-point of the
interval in which the first level change is detected, and $q_l$ be the
level crossed. 

From Fact~\ref{fact:samplepaths} and Remark~\ref{rem:techpoint1},
almost-\-surely the random sampling set $\mathcal{T}$ satisfies the
conditions of Fact~\ref{fact:zoransmainthm}. In addition, from
Fact~\ref{fact:samplepaths} and Proposition~\ref{prop:bddslope},
almost-\-surely, all sample paths of $X$ are differentiable
everywhere, and the derivative $X'$ has a normalized amplitude which
is uniformly bounded by $\pi$.  Hence, almost-\-surely for all paths of
$X$, the expansion (\ref{eq:samprep2}) holds, and by
(\ref{eq:precisionofsamples}),
\begin{eqnarray}
\label{eq:stocsampRDbbitdith}
\left|X(T_l)-\left(q_l-d_b\left(T_l\right)\right)\right| &\leq&
 \left(\frac{\pi + \widetilde{\Delta}}{\lambda}\right)
 2^{-\frac{1}{\lambda}R},
\end{eqnarray}
$\forall l\in\zZ$.  For each $t\in\rR$, let the bitrate-$R$, $b$-bit
quantized reconstruction of $X$ be defined as:
\begin{eqnarray}
\widehat{X}^{\mbox{\tiny $b$-bit}}_R(t) &:=&
\sum_{l\in\zZ}\left(q_l-d_b\left(T_l\right)\right)\psi_l\left(t-T_l\right),
\label{eq:defstocbbitdithrec}
\end{eqnarray}
where for the same reasons as for (\ref{eq:samprep2}), almost-\-surely
and in the expected $p$-th power sense, $p \in (0,\infty)$, the above
series converges absolutely for each $t$ and uniformly on all compact
subsets of $\rR$.  Thus as in (\ref{eq:bbitdetRDprofile}),
almost-\-surely,
\begin{eqnarray}
D(R) = ||X - \widehat{X}^{{\mbox{\tiny $b$-bit}}}_R||_\infty 
&\leq& \widetilde{C}'' 2^{-\frac{1}{\lambda}R},
\label{eq:bbitstocRDprofile}
\end{eqnarray}
with $\widetilde{C}'' := (\widetilde{C}'(\pi +
\widetilde{\Delta})/\lambda)$. This also implies that
\begin{eqnarray*}
||(\eE|X - \widehat{X}^{\mbox{\tiny
  $b$-bit}}_{R}|^p)^{1/p}||_\infty&\leq& \widetilde{C}''
  2^{-\frac{1}{\lambda}R},
\end{eqnarray*}
for all $p\in(0,\infty)$.  Once again, as discussed in
Section~\ref{sec:distcrit}, the bound $\widetilde{C}''
2^{-\frac{1}{\lambda}R}$ in (\ref{eq:bbitdetRDprofile}) and
(\ref{eq:bbitstocRDprofile}) also applies to the locally-averaged
reconstruction errors. It should be noted that $\widetilde{C}'$ only
depends on $\lambda, \underline{\kappa}$, and $\bar{\kappa}$, which
are fixed, and not on the specific realization of the set of random
sampling locations $\mathcal{T}$ which satisfy the assumptions of
Fact~\ref{fact:zoransmainthm}. However, the set of interpolation
kernels $\{\psi_l\}_{l\in\zZ}$ depends, in general, on the {\em
random} sampling set $\mathcal{T}$. \\

{
\subsection{Discussion}

\noindent{\bf Nyquist-sampling as a special case:}} Note that the
Nyquist-sampling scheme is also subsumed by the proposed generalized
dithered sampling framework. Indeed, for $b=k$, the described
framework suggests using two, $k$-bit ADCs at locations
$\left(\frac{l}{\lambda}\right)$ and $\left(\frac{l}{\lambda}\right) +
\tau$. However, the dithered field is guaranteed to have a
level-crossing in $\left[\frac{l}{\lambda}, \frac{l}{\lambda} +
\tau\right)$. Hence, the second sensor is redundant and there is no
need for the first sensor to add the dither-value. \\

\noindent{\bf Number of sensors vs ADC-precision and sensor
distribution:} While maintaining the same asymptotic error decay
profile, for a given reconstruction quality, as the ADC-precision $b$
increases, the number of sensors decreases exponentially with $b$.
One-bit dithered sampling needs $N$ sensors, $b$-bit ($b>1$) dithered
sampling needs $N_b=N/2^{b-1}$ sensors, and Nyquist-sampling needs
only one sensor per Nyquist-interval. The number of sensors can be
traded off with the precision of sensors. Although the number of
sensors has been a power two in our analysis, this is only for ease of
illustration and is not a restrictive assumption. Also, sensors need
to be placed only in intervals of the form $\left[\frac{l}{\lambda},
\frac{l}{\lambda}+(N_b-1)\tau \right)$. This leaves ``inactive''
regions over which the field is not sampled. Although the discussion
has, thus far, focused on sampling all the Nyquist-intervals in an
identical fashion, this need not be so. Some Nyquist-intervals can use
fewer sensors of higher precision ($b$ large) bunched closer together
in a smaller interval. Others can use more sensors of lower precision
($b$ small) spread over a larger interval.  The key result which
enables such a general non-uniform sampling using heterogeneous
sensors while achieving the same exponential rate-distortion decay
characteristics is Fact~\ref{fact:zoransmainthm}.  This bunched
irregular sampling ability provides considerable flexibility in
selecting spatially adaptive sensor precision and deployment patterns.
For example, in difficult terrain or in the presence of occluding
obstacles, where sampling density is forced to be low, one can use
high-precision sensors, and use low-precision sensors, in terrain
where the sampling density can be high.

Two other side-benefits of the dither-based oversampling scheme worth
mentioning (but not quantified here) include (i)
a certain degree of natural robustness to node failures: if every
alternate node fails, the effective inter-node separation increases to
$2\tau$ and is equivalent to halving the precision of the ADCs 
and (ii) some measure of data security against eavesdropping: the
dither fields can be kept covert and there is considerable flexibility
in their design for a given reconstruction quality.

Table~\ref{tab:summarysampschemes} summarizes all the salient features
of different sampling schemes considered in this paper.
\begin{table*}[!htb]
\begin{center}
\caption{Comparison of different sampling schemes.}
\label{tab:summarysampschemes}
\begin{tabular}{|c||c|c|c|}\hline
Sampling scheme: & Nyquist & one-bit dither & $b$-bit dither, $1<b<k$\\
\hline\hline
{Sensor-precision (bits)}& $k$-bit & one-bit & $b$-bit \\ \hline
Sensors per NQ intvl. & 1  &  $2^k$ &  $2^{k-b+1}$ \\ \hline
Inter-sensor spacing & $\left(\frac{1}{\lambda}\right)$ & $\tau := \frac{1}{\lambda
  2^k}$ & $\tau := \frac{1}{\lambda 2^k}$ \\  \hline
 &  & $2^k$ sensors uniformly& $2^{k-b+1}$ sensors uniformly\\
Sensor distribution & 1 sensor at every $\left(\frac{l}{\lambda}\right)$ &spaced
by $\tau$ &spaced by $\tau$ over \\
& & over every $\left[\frac{l}{\lambda}, \frac{l+1}{\lambda} \right)$ & 
every $\left[\frac{l}{\lambda}, \frac{l}{\lambda}+\frac{1}{\lambda
2^{b-1}}\right)$\\ \hline
Bits per NQ intvl. (bits/meter)& $R = k\lambda$ & $R = k\lambda$ & $R =
(k+1)\lambda $ \\ \hline
{\bf Distortion-Rate} 
& $D = O\left(2^{-\frac{1}{\lambda}\cdot R}\right)$ & $D =  O\left(2^{-
\frac{1}{\lambda}\cdot R}\right)$& $D = O\left(2^{-
  \frac{1}{\lambda}\cdot R}\right)$ \\ \hline
Local comm.~cost bit-meters per NQ intvl. & 0 & $\leq \frac{2}{\lambda}$ & $\leq \left(\frac{b+1}{\lambda
2^{b-1}}\right)\leq \frac{2}{\lambda}$ \\  \hline
\end{tabular}
\end{center}
\end{table*}

\Section{
{ Distributed Processing and Communication Costs}}
\label{sec:sensornetworks}

{Sections~\ref{sec:NQintro}--\ref{sec:bcp} focused on the
quantization and interpolation aspects of sampling. In this section we
focus on distributed processing and communication aspects of dithered
oversampling for reconstructing a sequence of fields at a CPU in a
sensor network.  We first discuss the problem of field acquisition and
coding (Section~\ref{sec:DataAcqu}).  We describe two approaches for
processing and coding raw sensor data, one based on local
nearest-neighbor communication and the other on distributed source
coding, and derive per Nyquist-interval per snapshot costs. Since
spatially bandlimited fields are not spatially limited, it is
impractical to move data from an infinite number of spatial
Nyquist-intervals to the CPU. We address this issue by considering how
the field reconstruction error in a compact region of interest decays
as the field is sampled over a growing neighborhood of sensor
deployment (Section~\ref{sec:ITCPU}). Here we focus on multihop
communication to move all the processed and coded data from each
spatial Nyquist-interval to the CPU.  We also indicate, however, how a
Gaussian multiple access uplink may be used for the same purpose.

To simplify the presentation, we focus on one-bit sensors and $1$-D
deterministic fields in $BL([-\pi,\pi])$ and assume that the dither
field is such that in each Nyquist-interval (i) there is exactly one
zero-crossing (see Remark~\ref{rem:techpoint2}) and (ii) the
zero-crossing direction (the sign change from say, plus to minus,
moving from left to right) is fixed.  In Section~\ref{sec:2Dextn} we
briefly indicate how results can be easily extended to two and higher
dimensional spatially bandlimited fields by reusing the results for
$1$-D fields.}

\subsection{Field acquisition and coding}\label{sec:DataAcqu}

Let $N$, one-bit sensors be placed uniformly at every $\tau =
\frac{1}{\lambda N}$ meters in every Nyquist-interval
$[\frac{l}{\lambda}, \frac{(l+1)}{\lambda} - \tau ]$.  Periodically,
the sensors synchronously take snapshots of the 1-D spatially
bandlimited random field by comparing the field-value to the
dither-value at their respective locations. The dither-values are
assumed to be pre-stored during sensor deployment. Sensor clocks are
assumed to be synchronized during deployment. The temporal sequence of
the signs of the field plus dither-values constitutes the {\em raw}
data produced at each sensor.

We consider two approaches for processing and coding the raw data
produced at each sensor for communication to a 
CPU where the field snapshots are to be reconstructed. The first
approach involves (a) neighbor-to-neighbor local communication within
each Nyquist-interval to locate the zero-crossings followed by (b)
coding of the zero-crossing data at each sensor. The second approach
does not require inter-sensor collaboration and is based on the
principles of distributed source coding due to Slepian and Wolf
\cite{slepianwnc73,covertei91} (see \cite{pradhanrd2003} for practical
code constructions). Distributed source coding exploits the {\em
correlation} present in the sensor data, e.g., there is exactly one
zero-crossing in each Nyquist-interval. Slepian and Wolf proved that
the compression-efficiency of non-collaborative coding, in terms of
the average rate, is equal to that of collaborative coding for
discrete-valued data. The probability of correctly decoding data which
has been compressed using a distributed source code tends to one as
the number of snapshots (coding blocklength) approaches infinity.

\subsubsection{Coding using local communication}

{\em (a) Locating zero-crossings.}  The sensors at the locations
$\{\frac{l}{\lambda}\}_{l\in\zZ}$ are designated as the ``leading''
nodes for initiating local communication (confined to
Nyquist-intervals) for determining the zero-crossing within each
spatial Nyquist-interval for each time snapshot.  Corresponding to
each temporal snapshot of the field, the message passed by each node
to its neighbor (the right neighbor for definiteness), indicates two
things: (i) whether or not a zero-crossing has already been found by
some preceding sensor and (ii) if a zero-crossing has not yet been
found, the sign of the field plus dither at its location. The local
communication terminates at the last (right-extreme) sensor in the
Nyquist-interval. In each Nyquist-interval, the first sensor that
detects a sign change between what its left neighbor reports and its
own reading records a ``1''. Other sensors record a ``0'' for that
snapshot. { Thus, at the {\em fixed} inter-sensor
communication rate of $2$ bits per snapshot, the raw one-bit sensor
readings can be converted to zero-crossing information.  Note that the
local communication can be done in one shot after aggregating the data
across many snapshots.}

{\em (b) Compression of zero-crossing data.} Each sensor encodes
(compresses) the zero-crossing information -- the temporal sequence of
zeros and ones where one indicates the presence of a zero-crossing and
a zero the absence. This compression can again be done in one of two
ways. The first approach, not discussed here (but see next
subsection), involves using distributed source coding by exploiting
the correlation in the zero-crossing information. However, an
alternative to distributed source coding involves each sensor
independently entropy-coding (e.g., run-length coding) its
zero-crossing data ignoring the inter-sensor correlation.
The compression efficiency of independent coding depends on the
specific underlying correlation structure.  Let $p_i \in [0,1], i =
1,\ldots,N$, denote the fraction of snapshots in which the
zero-crossings occur at the $i$-th sensor in a Nyquist-interval. Since
there is exactly one zero-crossing in each Nyquist-interval in each
snapshot, $\sum_{i=1}^N p_i = 1$. Sensor $i$ can compress its
zero-crossing information to the rate of $h_2(p_i)$ bits per snapshot,
where $h_2(p) = -p\log_2(p) - (1-p)\log_2(1-p), p\in[0,1]$, is the
binary entropy function \cite{covertei91}. The total bitrate in a
Nyquist-interval is given by $\sum_{i=1}^Nh_2(p_i) \leq N h_2(1/N)$
where the inequality is due to the (strict) concavity of $h_2(\cdot)$
and Jensen's inequality \cite{covertei91}.  This upper bound will be
attained if all sensors see the same number of zero-crossings
on the average. Thus the total number of bits per Nyquist-interval per
snapshot $R_{NQ}(N)$ for independent entropy-coding is not more than
$N \cdot h_2(1/N) \leq 1 + \log N$ bits per snapshot per
Nyquist-interval. Thus $R_{NQ}(N) = O(\log N)$.

\subsubsection{Distributed coding of raw data}\label{sec:dsc}
First, consider the {\em joint} compression of the raw binary data
gathered by all sensors in a Nyquist-interval. By design, there is
exactly one zero-crossing (whose direction is fixed) per
Nyquist-interval per snapshot. Hence, the $N$-length vector of raw
binary sensor data in any Nyquist-interval can take only $N$ distinct
values in each snapshot.  Hence, even while ignoring the exact
statistical structure of the temporal evolution, in the worst case, no
more than $\log_2(N)$ bits per snapshot per Nyquist-interval are
needed to encode these $N$-length vectors (being the maximum
entropy-rate of a source over an alphabet of size $N$
\cite{covertei91}). This represents the worst-case {\em joint
entropy-rate} (whenever the entropy rate exists) of all the
(binary-valued) sensor outputs in any Nyquist-interval.  However,
there exist distributed source codes
\cite{slepianwnc73,covertei91,pradhanrd2003} which, by exploiting the
underlying statistical correlation between sensor outputs, can achieve
a compression efficiency of $R_{NQ}(N) = \log_2(N)$ bits per
Nyquist-interval per snapshot {\em without any sensor collaboration}.
Further, if the joint statistics of all binary sensor data within each
Nyquist-interval has a symmetric structure,
then it is possible to noncollaboratively encode the data at each
sensor at the {\em evenly distributed} per-sensor rate of
$R_{sensor}(N) = \frac{1}{N}\log_2(N)$ bits per snapshot.

\subsection{Information transport to CPU and Scaling Law} \label{sec:ITCPU}

{ If the processed and coded data from each
Nyquist-interval from each snapshot can be somehow moved to the CPU
then the distortion (maximum expected point error) in reconstructing
each temporal snapshot will goes to zero as $D = O(2^{-R_{NQ}}) =
O(\frac{1}{N})$. Equivalently, for a fixed $D > 0$, the number of
one-bit sensors per Nyquist-interval needed will be finite with $N
\propto \frac{1}{D}$. If the field is also temporally bandlimited and
snapshots are taken at a stable temporal Nyquist-sampling rate then
each temporal slice can be reconstructed to a precision
$O(\frac{1}{N})$.\footnote{Every lower-dimensional spatio-temporal
slice of a bandlimited field is bandlimited to the same set of
frequencies.} Hence, through temporal interpolation, the entire
spatio-temporal field, that is, the field value at each point and at
each time instant, can be reconstructed to a point-wise precision of
$O(\frac{1}{N})$. 
Spatially bandlimited fields, however, are not spatially limited:
there are an infinite number of Nyquist-intervals. A practical way to
address this problem is to consider how the field reconstruction error
in a compact region of interest decays as the field is sampled over a
growing neighborhood of sensor deployment.}

\begin{figure}[!htb]
\begin{center}
\scalebox{0.75}{\input{./Figures/finite_region.pstex_t}}
\end{center}
\caption{\label{fig:finite_region} \sl \small 
The field is reconstructed over $[-L/\lambda,L/\lambda]$ by sampling
over a larger region $[-(L+L')/\lambda,(L+L')/\lambda]$ to meet a
desired reconstruction quality.  Distortion $D(N)$, geographical size
$\frac{2}{\lambda}(L+L'(N))$, and network rate $R_{net}(N)$ are
inter-related and depend on the number of one-bit sensors per
Nyquist-length $N$. Links closer to the CPU carry greater (but always
finite) traffic.}
\end{figure}
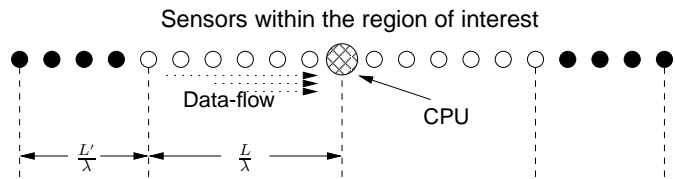

{
\subsubsection{Field reconstruction in a compact region of interest}

Let} $\left[-\frac{L}{\lambda}, \frac{L}{\lambda}\right]$ be the
interval of interest with the CPU located at the origin.  A symmetric
interval around the CPU is considered only for clarity and is not a
restrictive assumption.  The field is sampled in the larger interval
$\left[ - \frac{L+L'}{\lambda}, \frac{L+L'}{\lambda} \right]$.
{We want to analyze realizable rates of decay of
reconstruction error with $L'$.}  Over each Nyquist-interval, consider
$N$ one-bit sensors spread uniformly as in
Figure~\ref{fig:finite_region}.

For each time snapshot of the field, let $\mathcal{T} := \{t_l, l\in
\mathcal{I} \}$ be the collection of locations corresponding to the
mid-points of those inter-sensor intervals where the {\em first}
sign-change is detected in each Nyquist-interval. Here, $\mathcal{I}
:= \{-L-L', \ldots, L + L'\}$. As described in the previous
subsection,
the zero-crossings within each Nyquist-interval, can be located using
neighbor-to-neighbor local communication or alternatively, the raw
binary data at each sensor can be compressed using a distributed
source code.  Let $P_l(t, \mathcal{T})$ denote the Lagrange
interpolation polynomial for the $l$-th sampling location in
$\mathcal{T}$ \cite{marksi1990}. Specifically, $P_l(t, \mathcal{T}) =
\prod_{j\neq l}(t-t_j)(t_l-t_j)^{-1}$. Let
\begin{eqnarray*}
f_{\mbox{\tiny{Lag}}}(t) & := & \sum_{l\in \mathcal{I}} (-d(t_l))
P_l(t, \mathcal{T})
\end{eqnarray*}
be the Lagrange-interpolated reconstruction of $f(t)$ based on the
samples at $t_l \in \mathcal{T}$, where $d(t)$ is the dither field as
defined in Section~\ref{sec:ditherintro}. Thus
$f_{\mbox{\tiny{Lag}}}(t_l) = -d(t_l)$ for all $l$.  In
\cite{zorands2000} it was shown that $\forall t \in \left[
-\frac{L}{\lambda}, \frac{L}{\lambda} \right]$,
\begin{equation}
|f(t) - f_{\mbox{\tiny{Lag}}}(t)| \leq \sqrt{L'} \left(
\frac{\pi}{2\lambda}\right)^{2L'+1} +  \frac{\widetilde{C}}{N} {L'}^2,
\label{eq:lagacc1}
\end{equation}
if $\lambda > \frac{\pi}{2}$. The first error term in
(\ref{eq:lagacc1}) arises from the finiteness of the sampling window
and the second error term is due to quantization effects.

If $L'(N)$ is chosen such that
$\left(\frac{\pi}{2\lambda}\right)^{2L'+1} = \frac{1}{N}$, then $
L'(N) = \Theta\left(\log N \right)$ because $\lambda > \frac{\pi}{2}$.
From (\ref{eq:lagacc1}) it follows that for this choice of $L'(N)$,
\begin{eqnarray}
D(N) := \sup_{t\in [-\frac{L}{\lambda}, \frac{L}{\lambda}]}|f(t) -
f_{\mbox{\tiny{Lag}}}(t)| = O\left(\frac{(\log N)^2}{N}\right).
\label{eq:lagacc2} 
\end{eqnarray}

{
\subsubsection{Scaling Law for Multihopping} \label{sec:DataTran}

Now suppose that the compressed data from each sensor (from a large
number of snapshots) is routed through neighbor-to-neighbor local
communication in the direction of the CPU.\footnote{A detailed
description of the physical-layer and scheduling aspects of the local
communication are beyond the scope of this work.}} The node-id
(address) information will be part of the header of each data packet
sent to the CPU but will occupy only a vanishingly small fraction of
the total data rate for a sufficiently large number of snapshots. The
focus is on the sustainability of data {\em rates} {with
delays taken out of the picture.
The total number of bits per snapshot generated by the {\em entire}
network is given by}
\begin{eqnarray}
R_{net}(N) := 2(L + L'(N))R_{NQ}(N) = \Theta\left((\log N)^2\right),
\label{eq:Rnet}
\end{eqnarray}
where $R_{NQ}(N) = (\log N)$ is the number of bits per
Nyquist-interval per snapshot.
Eliminating $N$ from (\ref{eq:lagacc2}) and (\ref{eq:Rnet}) gives
\begin{eqnarray}
D(R_{net}) =  O\left( R_{net}2^{-\beta\sqrt{R_{net}}} \right),
\label{eq:DRnet}
\end{eqnarray}
where $\beta = \sqrt{\log \frac{2\lambda}{\pi} - o(1)}$. From
(\ref{eq:lagacc2}), (\ref{eq:Rnet}), and (\ref{eq:DRnet}) it follows
that any {\em nonzero} target distortion $D>0$ can be attained at the
CPU provided that $N$ (and $L'(N)$) are sufficiently large (but
finite) and the capacity of the busiest links (last-hop to the CPU) is
of the order of $R_{net}(N)$ bits per snapshot.  The network traffic
decreases approximately linearly with distance from the CPU.
{We summarize these results in terms of the following
realizable scaling law.

{\bf Multihop Scaling Law:} With $N$ one-bit sensors per
Nyquist-interval, $\Theta(\log N)$ Nyquist-intervals, and total
network bitrate $R_{net} = \Theta((\log N)^2)$ (per-sensor bitrate
$\Theta((\log N)/N)$), the maximum pointwise distortion goes to zero
as $D = O((\log N)^2/N)$ or $D = O(R_{net} 2^{-\beta
\sqrt{R_{net}}})$.}

\subsubsection{Gaussian Multiple Access Uplink} 
{ As an alternative to multihopping, suppose} that a real
Gaussian multiple access channel, bandlimited to $W_C$ Hertz, can be
realized for the effective multiuser uplink communication from all the
sensors {over $\Theta(\log N)$ Nyquist-intervals} to the
CPU. Then the {\em sum capacity}, in bits per second, of the uplink
channel grows as $C_{sum}(W_C,P_{tot}) =
W_C\log_2(1+\Theta(P_{tot}(N)))$ \cite{HanlyTseAllerton95}, where
$P_{tot}$ is the sum power of all {$\Theta(N\log N)$}
sensors. A sufficient condition, based on the separation of
distributed source coding and multiuser channel coding, for successful
transport of sensor data to the CPU is given by $C_{sum} > R_{NQ}
\times \mbox{snapshots per second}$ \cite{covertei91}. If $W_C > 1$
and each sensor has a sufficiently large (but fixed) power, then
$P_{tot} = \Theta(N{\log N})$, $C_{sum} > R_{NQ}(N) \times
\mbox{snapshots per second}$, and successful data-transport to the CPU
can be achieved {with distortion decreasing as in
(\ref{eq:lagacc2}).}

{
\subsubsection{Discussion} 

As} $N$ tends to infinity, $D$ approaches zero. However, for a fixed,
{\it nonzero} target distortion $D$, the number of sensors and the
network rate needed is always {\it finite}. This should be contrasted
with the following result discussed in \cite{marcodlnoIPSN2003} in an
information-theoretic setting: if all sensors use {\em identical}
deterministic scalar-quantizers then even if the sensor density
approaches infinity, the field reconstruction MSE is bounded strictly
away from zero. { The results in \cite{marcodlnoIPSN2003}
hold for all processes for which the probability of crossing some
quantizer threshold is positive. This includes bounded WSS bandlimited
processes studied in our work and also possibly those which do not
have a sufficient density of quantization threshold crossings for the
reconstruction error to converge to zero as the sensor density
increases to infinity \cite{ZoranV-OS:IT01}.  The use of identical
scalar quantizers can fundamentally limit the ability to decrease
distortion by increasing sensor density. This is clearly seen by
considering the extreme example of reconstructing a {\em constant}
field using identical one-bit scalar quantizers. This suggests the
need for having ``diversity'' in scalar quantization. Dithering is one
way of achieving this diversity.}

Also, as the desired distortion at the CPU approaches zero, one of two
things will {\em necessarily} have to take place to sustain the
desired performance: (i) If the CPU has only a finite number of links
then the capacity of at least one of these links must approach
infinity. This consequence is fundamental to rate-distortion theory
and cannot be bypassed because for any continuous source and any
reasonable distortion criterion, such as the space and time averaged
mean squared error (MSE), zero distortion requires infinite rate. (ii)
If the maximum capacity of each link is bounded then the number of
links (independent communication channels) needed to sustain the
desired level of reconstruction quality must go to infinity.

\subsection{Extension to two and higher spatial dimensions}\label{sec:2Dextn}

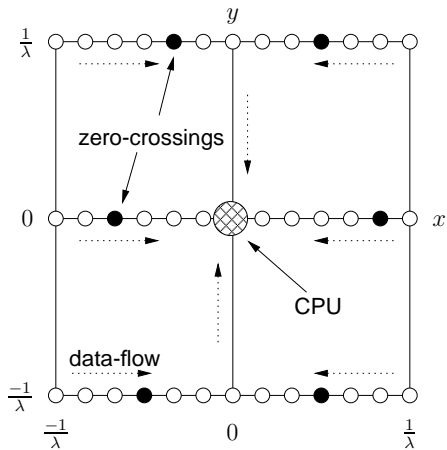
\begin{figure}[!htb]
\begin{center}
\scalebox{0.772}{\input{./Figures/twodim.pstex_t}}
\end{center}
\caption{\label{fig:twodim} \sl \small {\bf $2$-D distributed sampling
of spatially bandlimited fields.} Horizontal dimension densely
sampled: $N$ one-bit sensors uniformly distributed over each
horizontal Nyquist-interval. Vertical dimension sampled at the
vertical Nyquist-period $1/\lambda$.
Arrows indicate flow of compressed data towards the CPU. Distortion
$\downarrow 0$ as $O(1/N)$ as the rate per Nyquist-area $\uparrow$ as
$(\log N)$ like the 1-D case.}
\end{figure}
{ Although our focus has been on $1$-D fields, the} idea
of Nyquist space-slices can be used to effect a simple extension of
results { (though by no means the best possible)} from the
1-D one-bit dithered sampling framework {(with
multihopping)} to two and higher dimensions.  { We only
explain the } essence of this idea, {using}
Figure~\ref{fig:twodim}. {The key observation is that} it
is sufficient to sample only one out of the many orthogonal spatial
dimensions densely (and uniformly) using one-bit sensors while the
other spatial dimensions are sampled at their respective
Nyquist-rates. For example, for $2$-D spatially bandlimited fields,
one-bit sensors can be deployed along parallel horizontal lines where
the inter-line spacing is equal to the Nyquist-period along the
vertical dimension. With $N$ sensors spread uniformly over a
horizontal Nyquist-length (rather than over an area), the number of
sensors per Nyquist-area is also $N$.  The field along each horizontal
line corresponds to a horizontal spatial {\em slice} of the field and
is therefore bandlimited to the same set of horizontal frequencies.
Applying the results of the 1-D case it follows that the maximum
reconstruction error of each Nyquist space-slice (and hence the entire
spatial field) is of the order of $O(1/N)$. The bits per snapshot per
Nyquist-area grows as $(\log N)$.  If the field is also temporally
bandlimited, snapshots can be taken at the temporal Nyquist-rate.

\Section{Concluding remarks} \label{sec:conclusion}

The problem of deterministic and stochastic distributed sampling of
bandlimited sensor fields with low-precision ADCs was addressed in
this work. The possibility of achieving any desired reconstruction
quality using
a multitude of low-precision sensors was shown using a dither-based
sampling scheme. More importantly, the feasibility of flexible
tradeoffs between the oversampling rate and the ADC-precision with
respect to achieving exponential accuracy in the number of bits per
Nyquist-interval was demonstrated. This exposed a key underlying
conservation of bits principle.  
An achievable information scaling law for field reconstruction was
also derived.  Interestingly, this is possible using only
neighbor-to-neighbor communication and distributed source coding,
making it attractive for sensor networks. The extension of results
from the deterministic to the stochastic setting in a strong
almost-\-sure sense was also carried out. This paper is but a first
step towards understanding the fundamentals of distributed sampling
theory.  Interesting research directions include extensions to fields
with an unbounded dynamic range, e.g., Gaussian random fields, and
addressing the effects of sensing noise, random deployment,
location/timing errors and uncertainties, and synchronization and
scheduling issues.

\section*{Acknowledgment} The authors would like to thank
 Prof.~D.~L.~Neuhoff and Prof.~S.~S.~Pradhan (both EECS
Dept.~UMich.~Ann Arbor) for sharing their perspectives on distributed
field estimation in the sensor network context.

\appendices
\renewcommand{\theequation}{\thesection.\arabic{equation}}
\setcounter{equation}{0}

\Section{Examples of amplitude limited, bandlimited stationary
processes} \label{ap:WSSconstr}

Two examples of stationary processes with uniformly amplitude limited
sample paths and autocorrelation functions belonging to $BL([-W,W])$
for some $W > 0$ are presented.  The first one is strictly stationary,
the second is WSS.

\noindent \underline{Example~1:} Let $Z(t), t\in\rR$, be any strictly
stationary process (e.g., any WSS Gaussian process). By passing $Z$
through a hard-limiter (a pointwise operation), $y(z) := {\rm sign}
(z) \min(|z|,1)$, one obtains a strictly stationary process $Y(t),
t\in \rR $, which is amplitude limited. Let $X$ be the process
generated by passing $Y$ through any BIBO-stable\footnote{BIBO stands
for bounded-input-bounded-output.}  linear, shift-invariant (LSI)
system, with an impulse response belonging to
$BL([-W,W])\cap\mathcal{L}^1(\rR)$.  Then $X$ will (i) be amplitude
limited (in every sample path), being the output of a BIBO stable LSI
system with an amplitude limited input, (ii) stationary, being the
output of an LSI system with a stationary input, and (iii) have an
autocorrelation function belonging to $BL([-W,W])$, being the output
of a bandlimited LSI system.

\noindent \underline{Example~2:} Let $\{Y[l]\}_{l\in\zZ}$ be any
discrete, uniformly amplitude limited, WSS, white process, and
$\psi(t):= \left[\frac{\sin(\pi t)}{\pi t}\right]^2$ for $t\neq 0$ and
$\psi(0) := 1$.  Define a process
\begin{equation*}
X(t) :=   \lim_{L\longrightarrow\infty}\sum_{l=-L}^{L} Y[l] \psi(t + l + \Theta),\quad \forall t \in \rR,
\end{equation*}
where $\Theta \sim {\rm Uniform}[-1/2, 1/2]$ is independent of
$\{Y[l]\}_{l\in \zZ}$.  Note that $\psi$ is bandlimited, belongs to
$\mathcal{L}^1(\rR)\cap\mathcal{L}^2(\rR)$, and $\sup_{t\in\rR}
\left\{\sum_{l\in\zZ}|\psi(t+l)|\right\} = C < \infty$. Using this, it
can be shown that $X$ is a well defined WSS process, with a finite
mean, and an autocorrelation function which belongs to $BL([-W,W])$
and is uniformly amplitude limited.
Moreover, every sample path of $X$ is also uniformly amplitude
limited.

\Section{Proof of Proposition~\ref{prop:bddslope}} \label{ap:bddslope}

$h(t)$ is differentiable at each $t$, and the derivative $h' \in
\mathcal{L}^1(\rR)$ because $h'$ is uniformly amplitude limited and
decays as $\frac{1}{|t|^3}$.  Define $f'(t) := \int_{\rR} f(\tau)
h'(t-\tau)\mathrm{d}\tau$ and note that $|f'(t)| \leq A ||h'||_1 \leq
A(W+W+\delta)(W+\delta) < \infty$. { Since $f \in
ZBL(W,\delta)$ for all $\delta > 0$, we have $|f'(t)| \leq 2AW^2$.}

It remains to show that the derivative of $f(t)$ is $f'(t)$. To see
this, define $h_{\epsilon}'(t) :=
\frac{1}{\epsilon}\left(h(t+\epsilon) - h(t)\right), \epsilon \in
(0,1]$ and note that $|h_{\epsilon}'(t) - h'(t)|$ goes to zero with
decreasing $\epsilon$ for each $t\in\rR$ and can be bounded from above
by an absolutely integrable function $g(t)$ that does not depend on
$\epsilon$. Since $|\frac{1}{\epsilon} \left(f(t+\epsilon) -
f(t)\right) - f'(t)| \leq A ||h_{\epsilon}' - h'||_1$, the result
follows from Lebesgue's Dominated Convergence Theorem
\cite[Theorem~1.34]{RudinRealCmplx}. In fact, by using the same
technique recursively, it can be demonstrated that $f(t)$ is
differentiable an arbitrary number of times and all derivatives are
uniformly amplitude limited.
\hspace*{\fill}~\QED

\Section{Lifting sampling and interpolation results from $BL([-W,W])$
to $ZBL(W,\delta)$} \label{ap:Zakaiextn}

\noindent
For all $(t,\tau) \in \rR^2$, as $L \uparrow \infty$ along positive
integers,
\begin{eqnarray*}
\nu_L(t,\tau) &:=& \sum_{l=-L}^{L}
|f(\tau)||h(t_l-\tau)||\psi_l(t-t_l)| \quad \uparrow \\
\nu(t,\tau) &:=&
\sum_{l=-\infty}^{\infty} |f(\tau)||h(t_l-\tau)||\psi_l(t-t_l)|.
\end{eqnarray*}
Since $||f||_\infty, ||h||_\infty < \infty$ and for all $(t,L) \in
\rR\times \nN,~\sum_{l=-L}^{L}|\psi_l(t-t_l)| \leq C < \infty$,
\[
0 \leq \nu(t,\tau) \leq ||f||_\infty ||h||_\infty C < \infty.
\]
Also, since $||h||_1 < \infty$, for all $(t,L) \in \rR \times \nN$,
\[
\int_\rR \nu_L(t,\tau) \mathrm{d}\tau~\leq C ||f||_\infty ||h||_1 < \infty.
\]
To summarize the above in words, $\nu(t,\tau)$ is a nonnegative,
uniformly amplitude limited, measurable field and for all $(t,\tau)
\in \rR^2$, $\nu_L(t,\tau)$ is a nondecreasing sequence of nonnegative
real numbers converging to $\nu(t,\tau)$. For each $(t,L) \in \rR
\times \nN$, $\nu_L(t,\tau)$ is a measurable, absolutely-\-integrable
function of $\tau$. Hence, by Lebesgue's Monotone Convergence Theorem
\cite[Theorem~1.26]{RudinRealCmplx}, for all $t \in \rR$,
\[
\int_\rR \nu(t,\tau) \mathrm{d}\tau = \lim_{L\rightarrow \infty}
\int_\rR \nu_L(t,\tau) \mathrm{d}\tau \leq C ||f||_\infty ||h||_1.
\]
Thus, for all $t \in \rR$, $\nu(t,\tau)$ is a measurable,
absolutely-\-integrable function of $\tau$. For all $(t,\tau,L) \in
\rR^2\times \nN$ let
\[
g_L(t,\tau) := f(\tau) \sum_{l=-L}^L h(t_l-\tau)\psi_l(t-t_l).
\]
By assumption (ii) in the Lemma and because $||f||_\infty < \infty$,
as $L \uparrow \infty$ along positive integers, for all $(t,\tau) \in
\rR^2$,
\[
g_L(t,\tau) \rightarrow f(\tau)h(t-\tau).
\]
Furthermore, for all $(t,\tau) \in \rR^2$,
\[
|g_L(t,\tau)| \leq \nu(t,\tau),
\]
that is, for each $t\in\rR$, $g_L(t,\tau)$ is dominated by
$\nu(t,\tau)$ which is an absolutely-\-integrable function of
$\tau$. Hence, by Lebesgue's Dominated Convergence Theorem
\cite[Theorem~1.34]{RudinRealCmplx},
\begin{eqnarray}
\int_\rR f(\tau)h(t-\tau) \mathrm{d}\tau & = & \lim_{L\rightarrow\infty}
\int_\rR g_L(t,\tau) \mathrm{d}\tau, \label{eq:extmain}
\end{eqnarray}
for all $t \in \rR$. Since $f(\tau)$ is a Zakai-sense bandlimited
function of $\tau$, with associated Toeplitz reproducing kernel
$h(\tau)$, from (\ref{eq:reprodK}) we have for all $t \in \rR$,
\begin{eqnarray}
f(t) &=& \int_\rR f(\tau)h(t-\tau) \mathrm{d}\tau \label{eq:extdetails1}
\end{eqnarray}
and also
\begin{eqnarray}
\int_\rR g_L(t,\tau) \mathrm{d}\tau & = &
\sum_{l=-L}^L\psi_l(t-t_l)\int_\rR f(\tau) h(t_l-\tau) \mathrm{d}\tau \nonumber
\\
&=& \sum_{l=-L}^L f(t_l)\psi_l(t-t_l). \label{eq:extdetails2}
\end{eqnarray}
Hence, from (\ref{eq:extmain}), (\ref{eq:extdetails1}), and
(\ref{eq:extdetails2}), we have shown that for all $t\in\rR$,
\begin{eqnarray}
f(t) &=& \lim_{L\rightarrow \infty}\sum_{l=-L}^L
f(t_l)\psi_l(t-t_l). \label{eq:extfinal}
\end{eqnarray}
Since $||f||_\infty < \infty$ and for all $(t,L) \in \rR\times
\nN,~\sum_{l=-L}^{L}|\psi_l(t-t_l)| \leq C < \infty$, the series in
(\ref{eq:extfinal}) converges absolutely and uniformly on all compact
subsets of $\rR$. This completes the proof of the extension of
sampling and interpolation results from $BL([-W,W])$ to
$ZBL(W,\delta)$.
\hspace*{\fill}~\QED

\Section{Design of $b$-bit dither fields} \label{ap:bbitdither}

There are several possible approaches for designing a $b$-bit dither
field. Here we present only a simple natural construction based on
dither fields for one-bit sampling ($b=1$). For any $1 < b < k$, let
$M := 2^{b - 1}$.  For the design, it is sufficient to consider the
single interval $[A_0, B_0] = [0, \frac{1}{M\lambda} - \tau], \tau =
1/(2^k\lambda)$, i.e., the interval next to the origin corresponding
to $l = 0$ in Section~\ref{sec:bcp}. We define a $b$-bit dither field
as follows
\begin{eqnarray}
d_b(t) = \frac{2c}{M\gamma} d(t/(\lambda B_0)), \label{eq:bfrom1}
\end{eqnarray}
where $c > (1+\pi)/4$ is any constant and $d(t)$ is any dither field
used in the one-bit dithered sampling scheme of
Section~\ref{sec:ditherintro} with maximum amplitude $\gamma > 1$ and
a derivative whose amplitude is bounded by $\Delta$. From
Section~\ref{sec:onebitdithdet} we recall that $d(0) = -d(1/\lambda) =
\pm \gamma$.  We may assume, without loss of generality, that $d(0) >
0$ so that $d_b(A_0) = - d_b(B_0) = 2c/M > 2/M$.  Also note that $|d_b
'(t)| \leq \frac{2c}{M\gamma}\frac{1}{\lambda B_0} \Delta =
\frac{2c\Delta}{\gamma}(1- \frac{2^{b-1}}{2^k})^{-1} <
\frac{4c}{\gamma}\Delta$ because $1 < b < k$. Thus, $\widetilde{\Delta} := \sup
|d_b'(t)| < \frac{4c}{\gamma} \Delta$.

It will be shown that the dithered-field $[f + d_b](t)$ crosses a
level in the set $\{0, \pm \frac{1}{M}, \ldots, \pm \frac{M-1}{M} \}$
in the interval $[A_0, B_0]$. By design, $d_b(t)$ is continuous. The
intermediate value theorem for continuous functions is used to prove
the existence of a level-crossing. Specifically, it will be shown that
$[f+d_b](A_0)$ and $[f+d_b](B_0)$ lie on different sides of some
level. Let $L_j, -M \leq j \leq (M-1)$, be quantization intervals
given by
\begin{eqnarray*}
L_j = 
\begin{cases}
\left(-\infty,-\frac{M-1}{M} \right) & \mbox{ for } j = -M, \\
\left[\frac{j}{M}, \frac{j + 1 }{M}\right) & \mbox{ for } -(M-1) \leq
j \leq (M-2), \nonumber \\
\left[\frac{M-1}{M}, \infty \right) & \mbox{ for } j = (M-1). \nonumber \\
\end{cases}
\end{eqnarray*}
Since $d_b(A_0) > \frac{2}{M}$, therefore, $f(A_0) + d_b(A_0) > -1 +
\frac{2}{M}$.  Thus, $[f + d_b](A_0) \notin L_j$ for $j \leq -(M -
1)$.  So let $[f+ d_b](A_0) \in L_j$ for some $j$, $-(M - 2) \leq j
\leq (M -2)$. We now show that $[f+d_b](B_0)$ lies in a different
quantization interval than $L_j$. We have the following inequalities
\begin{eqnarray}
[f + d_b](B_0) & = & f(B_0) - \frac{2c}{M}, \nonumber \\ 
& \stackrel{(i)}{\leq} & f(A_0) + \pi(B_0-A_0) -
\frac{2c}{M},\nonumber \\
& \stackrel{(ii)}{<} & \frac{j+1}{M} - \frac{2c}{M} + \pi(B_0-A_0)
- \frac{2c}{M}, \nonumber \\
& \stackrel{(iii)}{\leq} & \frac{j}{M}, \nonumber
\end{eqnarray}
where $(i)$ follows from the mean value theorem applied to $f(t)$ at
the end points $A_0$ and $B_0$ and noting that $|f'(t)|$ is bounded by
$\pi$ (Bernstein's inequality), $(ii)$ is because $[f+d_b](A_0) =
f(A_0) + \frac{2c}{M} \in L_j, -(M - 2) \leq j \leq (M -2),
\Rightarrow f(A_0) + \frac{2c}{M} < \frac{(j + 1)}{M}$, and $(iii)$
follows from $c > (1+\pi)/4$, $\lambda > 1$, and some algebra.  Since
$[f + d_b](B_0) < j/M$,
$[f + d_b](B_0) \in L_{j'}$ for some $j' \leq (j-1)$, i.e., $[f +
d_b](B_0) \notin L_{j}$.  This together with the assumption that $[f +
d_b](A_0) \in L_j, -(M - 2) \leq j \leq (M -2)$ shows that there is a
level-crossing of $[f + d_b](t)$ in the interval $[A_0, B_0]$.

In the special case when $[f + d_b](A_0) \in L_{(M-1)}$, then $[f +
d_b](B_0) = f(B_0) - 2c/M < 1 - 2/M$. Hence, $[f + d_b](B_0)$ lies in
the quantization interval $L_i, i \leq (M-3)$ and the level $(M -
1)/M$ will be crossed by the dithered field.

\bibliography{references}

\end{document}

%% file: Figures/Zakai.pstex_t
\begin{picture}(0,0)%
\includegraphics{Figures/Zakai.pstex}%
\end{picture}%
\setlength{\unitlength}{3947sp}%
\begingroup\makeatletter\ifx\SetFigFont\undefined%
\gdef\SetFigFont#1#2#3#4#5{%
  \reset@font\fontsize{#1}{#2pt}%
  \fontfamily{#3}\fontseries{#4}\fontshape{#5}%
  \selectfont}%
\fi\endgroup%
\begin{picture}(5722,3253)(3879,-5092)
\put(7651,-2161){\makebox(0,0)[b]{\smash{{\SetFigFont{20}{24.0}{\familydefault}{\mddefault}{\updefault}{\color[rgb]{0,0,0}$H(\omega) \leftrightarrow h(t)$}%
}}}}
\put(6751,-4186){\makebox(0,0)[b]{\smash{{\SetFigFont{20}{24.0}{\familydefault}{\mddefault}{\updefault}{\color[rgb]{0,0,0}$0$}%
}}}}
\put(9601,-4261){\makebox(0,0)[b]{\smash{{\SetFigFont{20}{24.0}{\familydefault}{\mddefault}{\updefault}{\color[rgb]{0,0,0}$\omega$}%
}}}}
\put(7801,-4636){\makebox(0,0)[b]{\smash{{\SetFigFont{20}{24.0}{\familydefault}{\mddefault}{\updefault}{\color[rgb]{0,0,0}$W$}%
}}}}
\put(5401,-4636){\makebox(0,0)[b]{\smash{{\SetFigFont{20}{24.0}{\familydefault}{\mddefault}{\updefault}{\color[rgb]{0,0,0}$-W$}%
}}}}
\put(8101,-5011){\makebox(0,0)[b]{\smash{{\SetFigFont{20}{24.0}{\familydefault}{\mddefault}{\updefault}{\color[rgb]{0,0,0}$\delta$}%
}}}}
\end{picture}%

%% file: Figures/classicNQ.pstex_t
\begin{picture}(0,0)%
\includegraphics{./Figures/classicNQ.pstex}%
\end{picture}%
\setlength{\unitlength}{3158sp}%
\begingroup\makeatletter\ifx\SetFigFont\undefined%
\gdef\SetFigFont#1#2#3#4#5{%
  \reset@font\fontsize{#1}{#2pt}%
  \fontfamily{#3}\fontseries{#4}\fontshape{#5}%
  \selectfont}%
\fi\endgroup%
\begin{picture}(5092,2811)(-58,-2017)
\put(226, 89){\makebox(0,0)[b]{\smash{{\SetFigFont{11}{13.2}{\rmdefault}{\mddefault}{\updefault}{\color[rgb]{0,0,0}$1$}%
}}}}
\put(2476,314){\makebox(0,0)[lb]{\smash{{\SetFigFont{12}{14.4}{\rmdefault}{\mddefault}{\updefault}{\color[rgb]{0,0,0}$f(t)$}%
}}}}
\put(2701,-1936){\makebox(0,0)[b]{\smash{{\SetFigFont{14}{16.8}{\rmdefault}{\mddefault}{\updefault}{\color[rgb]{0,0,0}$\frac{l+1}{\lambda}$}%
}}}}
\put(4951,-1936){\makebox(0,0)[b]{\smash{{\SetFigFont{14}{16.8}{\rmdefault}{\mddefault}{\updefault}{\color[rgb]{0,0,0}$\frac{l+2}{\lambda}$}%
}}}}
\put(751,-1936){\makebox(0,0)[b]{\smash{{\SetFigFont{14}{16.8}{\rmdefault}{\mddefault}{\updefault}{\color[rgb]{0,0,0}$\frac{l}{\lambda}$}%
}}}}
\put(226,-811){\makebox(0,0)[b]{\smash{{\SetFigFont{11}{13.2}{\rmdefault}{\mddefault}{\updefault}{\color[rgb]{0,0,0}$0$}%
}}}}
\put(226,-136){\makebox(0,0)[b]{\smash{{\SetFigFont{11}{13.2}{\rmdefault}{\mddefault}{\updefault}{\color[rgb]{0,0,0}$3/4$}%
}}}}
\put(226,-361){\makebox(0,0)[b]{\smash{{\SetFigFont{11}{13.2}{\rmdefault}{\mddefault}{\updefault}{\color[rgb]{0,0,0}$1/2$}%
}}}}
\put(226,-586){\makebox(0,0)[b]{\smash{{\SetFigFont{11}{13.2}{\rmdefault}{\mddefault}{\updefault}{\color[rgb]{0,0,0}$1/4$}%
}}}}
\put(226,-1261){\makebox(0,0)[b]{\smash{{\SetFigFont{11}{13.2}{\rmdefault}{\mddefault}{\updefault}{\color[rgb]{0,0,0}$-1/2$}%
}}}}
\put(226,-1486){\makebox(0,0)[b]{\smash{{\SetFigFont{11}{13.2}{\rmdefault}{\mddefault}{\updefault}{\color[rgb]{0,0,0}$-3/4$}%
}}}}
\put(226,-1711){\makebox(0,0)[b]{\smash{{\SetFigFont{11}{13.2}{\rmdefault}{\mddefault}{\updefault}{\color[rgb]{0,0,0}$-1$}%
}}}}
\put(226,-1036){\makebox(0,0)[b]{\smash{{\SetFigFont{11}{13.2}{\rmdefault}{\mddefault}{\updefault}{\color[rgb]{0,0,0}$-1/4$}%
}}}}
\end{picture}%

%% file: Figures/1bitdither.pstex_t
\begin{picture}(0,0)%
\includegraphics{Figures/1bitdither.pstex}%
\end{picture}%
\setlength{\unitlength}{3947sp}%
\begingroup\makeatletter\ifx\SetFigFont\undefined%
\gdef\SetFigFont#1#2#3#4#5{%
  \reset@font\fontsize{#1}{#2pt}%
  \fontfamily{#3}\fontseries{#4}\fontshape{#5}%
  \selectfont}%
\fi\endgroup%
\begin{picture}(7154,4185)(1516,-4342)
\put(6151,-1861){\makebox(0,0)[b]{\smash{{\SetFigFont{17}{20.4}{\familydefault}{\mddefault}{\updefault}{\color[rgb]{0,0,0}$2^k$ \textsf{sensors}, $k=3$}%
}}}}
\put(2776,-2086){\makebox(0,0)[b]{\smash{{\SetFigFont{17}{20.4}{\familydefault}{\mddefault}{\updefault}{\color[rgb]{0,0,0}$\tau$}%
}}}}
\put(5401,-361){\makebox(0,0)[b]{\smash{{\SetFigFont{17}{20.4}{\rmdefault}{\bfdefault}{\updefault}{\color[rgb]{0,0,0}Oversampled 1-bit dither}%
}}}}
\put(1801,-3661){\makebox(0,0)[b]{\smash{{\SetFigFont{17}{20.4}{\familydefault}{\mddefault}{\updefault}{\color[rgb]{0,0,0}$-1$}%
}}}}
\put(1801,-2536){\makebox(0,0)[b]{\smash{{\SetFigFont{17}{20.4}{\familydefault}{\mddefault}{\updefault}{\color[rgb]{0,0,0}$0$}%
}}}}
\put(1801,-1261){\makebox(0,0)[b]{\smash{{\SetFigFont{17}{20.4}{\familydefault}{\mddefault}{\updefault}{\color[rgb]{0,0,0}$+1$}%
}}}}
\put(5401,-961){\makebox(0,0)[b]{\smash{{\SetFigFont{17}{20.4}{\familydefault}{\mddefault}{\updefault}{\color[rgb]{0,0,0}$f(t) + d(t)$}%
}}}}
\put(4051,-3211){\makebox(0,0)[b]{\smash{{\SetFigFont{17}{20.4}{\familydefault}{\mddefault}{\updefault}{\color[rgb]{0,0,0}$z_l$}%
}}}}
\put(7276,-3211){\makebox(0,0)[b]{\smash{{\SetFigFont{17}{20.4}{\familydefault}{\mddefault}{\updefault}{\color[rgb]{0,0,0}$z_{l+1}$}%
}}}}
\put(2176,-4261){\makebox(0,0)[b]{\smash{{\SetFigFont{17}{20.4}{\familydefault}{\mddefault}{\updefault}{\color[rgb]{0,0,0}$\frac{l}{\lambda}$}%
}}}}
\put(5326,-4261){\makebox(0,0)[b]{\smash{{\SetFigFont{17}{20.4}{\familydefault}{\mddefault}{\updefault}{\color[rgb]{0,0,0}$\frac{l+1}{\lambda}$}%
}}}}
\put(8551,-4261){\makebox(0,0)[b]{\smash{{\SetFigFont{17}{20.4}{\familydefault}{\mddefault}{\updefault}{\color[rgb]{0,0,0}$\frac{l+2}{\lambda}$}%
}}}}
\end{picture}%

%% file: Figures/bcp.pstex_t
\begin{picture}(0,0)%
\includegraphics{./Figures/bcp.pstex}%
\end{picture}%
\setlength{\unitlength}{3158sp}%
\begingroup\makeatletter\ifx\SetFigFont\undefined%
\gdef\SetFigFont#1#2#3#4#5{%
  \reset@font\fontsize{#1}{#2pt}%
  \fontfamily{#3}\fontseries{#4}\fontshape{#5}%
  \selectfont}%
\fi\endgroup%
\begin{picture}(9644,3064)(-35,-2294)
\put(226,-811){\makebox(0,0)[b]{\smash{{\SetFigFont{10}{12.0}{\rmdefault}{\mddefault}{\updefault}{\color[rgb]{0,0,0}$0$}%
}}}}
\put(6151,-1936){\makebox(0,0)[b]{\smash{{\SetFigFont{14}{16.8}{\rmdefault}{\mddefault}{\updefault}{\color[rgb]{0,0,0}$\frac{l+1}{\lambda}$}%
}}}}
\put(676,-1936){\makebox(0,0)[b]{\smash{{\SetFigFont{14}{16.8}{\rmdefault}{\mddefault}{\updefault}{\color[rgb]{0,0,0}$\frac{l}{\lambda}$}%
}}}}
\put(3826,-1936){\makebox(0,0)[b]{\smash{{\SetFigFont{14}{16.8}{\rmdefault}{\mddefault}{\updefault}{\color[rgb]{0,0,0}$\frac{l}{\lambda}$}%
}}}}
\put(7276,-1936){\makebox(0,0)[b]{\smash{{\SetFigFont{14}{16.8}{\rmdefault}{\mddefault}{\updefault}{\color[rgb]{0,0,0}$\frac{l}{\lambda}$}%
}}}}
\put(7726,-1411){\makebox(0,0)[b]{\smash{{\SetFigFont{10}{12.0}{\familydefault}{\mddefault}{\updefault}{\color[rgb]{0,0,0}$z_l$}%
}}}}
\put(6826,-1711){\makebox(0,0)[b]{\smash{{\SetFigFont{10}{12.0}{\familydefault}{\mddefault}{\updefault}{\color[rgb]{0,0,0}$-1$}%
}}}}
\put(6826,-1261){\makebox(0,0)[b]{\smash{{\SetFigFont{10}{12.0}{\rmdefault}{\mddefault}{\updefault}{\color[rgb]{0,0,0}$-1/2$}%
}}}}
\put(6826,-811){\makebox(0,0)[b]{\smash{{\SetFigFont{10}{12.0}{\familydefault}{\mddefault}{\updefault}{\color[rgb]{0,0,0}$0$}%
}}}}
\put(6826,-361){\makebox(0,0)[b]{\smash{{\SetFigFont{10}{12.0}{\rmdefault}{\mddefault}{\updefault}{\color[rgb]{0,0,0}$1/2$}%
}}}}
\put(3451,-811){\makebox(0,0)[b]{\smash{{\SetFigFont{10}{12.0}{\familydefault}{\mddefault}{\updefault}{\color[rgb]{0,0,0}$0$}%
}}}}
\put(3451, 89){\makebox(0,0)[b]{\smash{{\SetFigFont{10}{12.0}{\familydefault}{\mddefault}{\updefault}{\color[rgb]{0,0,0}$+1$}%
}}}}
\put(3451,-1711){\makebox(0,0)[b]{\smash{{\SetFigFont{10}{12.0}{\familydefault}{\mddefault}{\updefault}{\color[rgb]{0,0,0}$-1$}%
}}}}
\put(6826, 89){\makebox(0,0)[b]{\smash{{\SetFigFont{10}{12.0}{\familydefault}{\mddefault}{\updefault}{\color[rgb]{0,0,0}$+1$}%
}}}}
\put(226,-1036){\makebox(0,0)[b]{\smash{{\SetFigFont{10}{12.0}{\rmdefault}{\mddefault}{\updefault}{\color[rgb]{0,0,0}$-1/4$}%
}}}}
\put(7876,314){\makebox(0,0)[lb]{\smash{{\SetFigFont{11}{13.2}{\rmdefault}{\mddefault}{\updefault}{\color[rgb]{0,0,0}$f(t)+d_2(t)$}%
}}}}
\put(226,-1486){\makebox(0,0)[b]{\smash{{\SetFigFont{10}{12.0}{\rmdefault}{\mddefault}{\updefault}{\color[rgb]{0,0,0}$-3/4$}%
}}}}
\put(226,-1711){\makebox(0,0)[b]{\smash{{\SetFigFont{10}{12.0}{\rmdefault}{\mddefault}{\updefault}{\color[rgb]{0,0,0}$-1$}%
}}}}
\put(2701,-1936){\makebox(0,0)[b]{\smash{{\SetFigFont{14}{16.8}{\rmdefault}{\mddefault}{\updefault}{\color[rgb]{0,0,0}$\frac{l+1}{\lambda}$}%
}}}}
\put(9526,-1936){\makebox(0,0)[b]{\smash{{\SetFigFont{14}{16.8}{\rmdefault}{\mddefault}{\updefault}{\color[rgb]{0,0,0}$\frac{l+1}{\lambda}$}%
}}}}
\put(7576,-436){\makebox(0,0)[b]{\smash{{\SetFigFont{10}{12.0}{\familydefault}{\mddefault}{\updefault}{\color[rgb]{0,0,0}$\tau$}%
}}}}
\put(5141,-1273){\makebox(0,0)[b]{\smash{{\SetFigFont{10}{12.0}{\familydefault}{\mddefault}{\updefault}{\color[rgb]{0,0,0}$z_l$}%
}}}}
\put(4201,-436){\makebox(0,0)[b]{\smash{{\SetFigFont{10}{12.0}{\familydefault}{\mddefault}{\updefault}{\color[rgb]{0,0,0}$\tau$}%
}}}}
\put(4501,314){\makebox(0,0)[lb]{\smash{{\SetFigFont{11}{13.2}{\rmdefault}{\mddefault}{\updefault}{\color[rgb]{0,0,0}$f(t)+d(t)$}%
}}}}
\put(1351,314){\makebox(0,0)[lb]{\smash{{\SetFigFont{11}{13.2}{\rmdefault}{\mddefault}{\updefault}{\color[rgb]{0,0,0}$f(t)$}%
}}}}
\put(226, 89){\makebox(0,0)[b]{\smash{{\SetFigFont{10}{12.0}{\rmdefault}{\mddefault}{\updefault}{\color[rgb]{0,0,0}$+1$}%
}}}}
\put(226,-136){\makebox(0,0)[b]{\smash{{\SetFigFont{10}{12.0}{\rmdefault}{\mddefault}{\updefault}{\color[rgb]{0,0,0}$+3/4$}%
}}}}
\put(226,-361){\makebox(0,0)[b]{\smash{{\SetFigFont{10}{12.0}{\rmdefault}{\mddefault}{\updefault}{\color[rgb]{0,0,0}$+1/2$}%
}}}}
\put(226,-586){\makebox(0,0)[b]{\smash{{\SetFigFont{10}{12.0}{\rmdefault}{\mddefault}{\updefault}{\color[rgb]{0,0,0}$+1/4$}%
}}}}
\put(226,-1261){\makebox(0,0)[b]{\smash{{\SetFigFont{10}{12.0}{\rmdefault}{\mddefault}{\updefault}{\color[rgb]{0,0,0}$-1/2$}%
}}}}
\put(8401,-2236){\makebox(0,0)[b]{\smash{{\SetFigFont{10}{12.0}{\familydefault}{\mddefault}{\updefault}{\color[rgb]{0,0,0}\textsf{(c)} $2^{k-b+1}=4$ \textsf{two-bit sensors}}%
}}}}
\put(1501,-2236){\makebox(0,0)[b]{\smash{{\SetFigFont{10}{12.0}{\familydefault}{\mddefault}{\updefault}{\color[rgb]{0,0,0}\textsf{(a)} $1$ \textsf{three-bit sensor}}%
}}}}
\put(5026,-2236){\makebox(0,0)[b]{\smash{{\SetFigFont{10}{12.0}{\familydefault}{\mddefault}{\updefault}{\color[rgb]{0,0,0}\textsf{(b)} $2^k=8$ \textsf{one-bit sensors}}%
}}}}
\end{picture}%

%% file: Figures/finite_region.pstex_t
\begin{picture}(0,0)%
\includegraphics{./Figures/finite_region.pstex}%
\end{picture}%
\setlength{\unitlength}{3552sp}%
\begingroup\makeatletter\ifx\SetFigFont\undefined%
\gdef\SetFigFont#1#2#3#4#5{%
  \reset@font\fontsize{#1}{#2pt}%
  \fontfamily{#3}\fontseries{#4}\fontshape{#5}%
  \selectfont}%
\fi\endgroup%
\begin{picture}(6166,1593)(818,-1798)
\put(1426,-1636){\makebox(0,0)[lb]{\smash{{\SetFigFont{11}{13.2}{\rmdefault}{\mddefault}{\updefault}{\color[rgb]{0,0,0}$\frac{L'}{\lambda}$}%
}}}}
\put(2926,-1636){\makebox(0,0)[lb]{\smash{{\SetFigFont{11}{13.2}{\rmdefault}{\mddefault}{\updefault}{\color[rgb]{0,0,0}$\frac{L}{\lambda}$}%
}}}}
\put(4951,-1486){\makebox(0,0)[b]{\smash{{\SetFigFont{11}{13.2}{\familydefault}{\mddefault}{\updefault}{\color[rgb]{0,0,0}\textsf{}}%
}}}}
\put(4876,-1261){\makebox(0,0)[b]{\smash{{\SetFigFont{11}{13.2}{\familydefault}{\mddefault}{\updefault}{\color[rgb]{0,0,0}\textsf{CPU}}%
}}}}
\put(3976,-361){\makebox(0,0)[b]{\smash{{\SetFigFont{12}{14.4}{\rmdefault}{\mddefault}{\updefault}{\color[rgb]{0,0,0}\textsf{Sensors within the region of interest}}%
}}}}
\put(2851,-1111){\makebox(0,0)[b]{\smash{{\SetFigFont{11}{13.2}{\rmdefault}{\mddefault}{\updefault}{\color[rgb]{0,0,0}\textsf{Data-flow}}%
}}}}
\end{picture}%

%% file: Figures/twodim.pstex_t
\begin{picture}(0,0)%
\includegraphics{./Figures/twodim.pstex}%
\end{picture}%
\setlength{\unitlength}{3158sp}%
\begingroup\makeatletter\ifx\SetFigFont\undefined%
\gdef\SetFigFont#1#2#3#4#5{%
  \reset@font\fontsize{#1}{#2pt}%
  \fontfamily{#3}\fontseries{#4}\fontshape{#5}%
  \selectfont}%
\fi\endgroup%
\begin{picture}(4679,4534)(6,-3878)
\put(4201,-3811){\makebox(0,0)[b]{\smash{{\SetFigFont{12}{14.4}{\rmdefault}{\mddefault}{\updefault}{\color[rgb]{0,0,0}$\frac{1}{\lambda}$}%
}}}}
\put(1576,-811){\makebox(0,0)[b]{\smash{{\SetFigFont{11}{13.2}{\rmdefault}{\mddefault}{\updefault}{\color[rgb]{0,0,0}\textsf{zero-crossings}}%
}}}}
\put(4501,-1636){\makebox(0,0)[b]{\smash{{\SetFigFont{12}{14.4}{\rmdefault}{\mddefault}{\updefault}{\color[rgb]{0,0,0}$x$}%
}}}}
\put(2401,464){\makebox(0,0)[b]{\smash{{\SetFigFont{12}{14.4}{\rmdefault}{\mddefault}{\updefault}{\color[rgb]{0,0,0}$y$}%
}}}}
\put(376,-3436){\makebox(0,0)[rb]{\smash{{\SetFigFont{12}{14.4}{\rmdefault}{\mddefault}{\updefault}{\color[rgb]{0,0,0}$\frac{-1}{\lambda}$}%
}}}}
\put(376,-1636){\makebox(0,0)[rb]{\smash{{\SetFigFont{12}{14.4}{\rmdefault}{\mddefault}{\updefault}{\color[rgb]{0,0,0}$0$}%
}}}}
\put(376,164){\makebox(0,0)[rb]{\smash{{\SetFigFont{12}{14.4}{\rmdefault}{\mddefault}{\updefault}{\color[rgb]{0,0,0}$\frac{1}{\lambda}$}%
}}}}
\put(3301,-2536){\makebox(0,0)[b]{\smash{{\SetFigFont{11}{13.2}{\rmdefault}{\mddefault}{\updefault}{\color[rgb]{0,0,0}\textsf{CPU           }}%
}}}}
\put(601,-3811){\makebox(0,0)[b]{\smash{{\SetFigFont{12}{14.4}{\rmdefault}{\mddefault}{\updefault}{\color[rgb]{0,0,0}$\frac{-1}{\lambda}$}%
}}}}
\put(2401,-3811){\makebox(0,0)[b]{\smash{{\SetFigFont{12}{14.4}{\rmdefault}{\mddefault}{\updefault}{\color[rgb]{0,0,0}$0$}%
}}}}
\put(1201,-3061){\makebox(0,0)[b]{\smash{{\SetFigFont{11}{13.2}{\rmdefault}{\mddefault}{\updefault}{\color[rgb]{0,0,0}\textsf{data-flow}}%
}}}}
\end{picture}%